\documentclass[12pt]{article}

\usepackage[letterpaper,hmargin=1in,vmargin=1in]{geometry}
\usepackage{color}
\usepackage{graphicx,epstopdf,amsmath,amsfonts,amssymb,cite}
\usepackage[hyperindex,breaklinks]{hyperref}
\usepackage[small,compact]{titlesec} 

\def\lsim{\mathrel{\rlap{\lower4pt\hbox{\hskip1pt$\sim$}}
    \raise1pt\hbox{$<$}}}                % less than or approx. symbol
\def\gsim{\mathrel{\rlap{\lower4pt\hbox{\hskip1pt$\sim$}}
    \raise1pt\hbox{$>$}}}                % greater than or approx. symbol
\def\be{\begin{equation}}
\def\ee{\end{equation}}
\def\ba{\begin{eqnarray}}
\def\ea{\end{eqnarray}}
\def\ge{\mathrel{\raise.3ex\hbox{$>$\kern-.75em\lower1ex\hbox{$\sim$}}}}
\def\la{\mathrel{\raise.3ex\hbox{$<$\kern-.75em\lower1ex\hbox{$\sim$}}}}

\def\blfootnote{\xdef\@thefnmark{}\@footnotetext}

\def\simgt{\mathrel{\raise.3ex\hbox{$>$\kern-.75em\lower1ex\hbox{$\sim$}}}}
\def\simlt{\mathrel{\raise.3ex\hbox{$<$\kern-.75em\lower1ex\hbox{$\sim$}}}}

\newcommand{\nc}{\newcommand}

\nc{\gone}{\bar g_{\pi NN}^{(1)}}
\nc{\gzero}{\bar g_{\pi NN}^{(0)}}
\nc{\al}{\alpha}
\nc{\ga}{\gamma}
\nc{\de}{\delta}
\nc{\ep}{\epsilon}
\nc{\ze}{\zeta}
\nc{\et}{\eta}
\nc{\ka}{\kappa}
\nc{\rh}{\rho}
\nc{\si}{\sigma}
\nc{\ta}{\tau}
\nc{\up}{\upsilon}
\nc{\ph}{\phi}
\nc{\ch}{\chi}
\nc{\ps}{\psi}
\nc{\om}{\omega}
\nc{\Ga}{\Gamma}
\nc{\De}{\Delta}
\nc{\La}{\Lambda}
\nc{\Si}{\Sigma}
\nc{\Up}{\Upsilon}
\nc{\Ph}{\Phi}
\nc{\Ps}{\Psi}
\nc{\Om}{\Omega}
\nc{\ptl}{\partial}
\nc{\del}{\nabla}
\nc{\ov}{\overline}
\nc{\newcaption}[1]{\centerline{\parbox{15cm}{\caption{#1}}}}
\nc{\us}{U(1)$_S$}

\def\beq{\begin{equation}}
\def\eeq{\end{equation}}
\def\bmat{\begin{displaymath}}
\def\emat{\end{displaymath}}
\def\bear{\begin{eqnarray}}
\def\eear{\end{eqnarray}}
\def\ba{\begin{eqnarray}}
\def\ea{\end{eqnarray}}
\def\bery{\begin{array}}
\def\ery{\end{array}}
\def\bit{\begin{itemize}}
\def\eit{\end{itemize}}
\def\ben{\begin{enumerate}}
\def\een{\end{enumerate}}
\def\btab{\begin{tabular}}
\def\etab{\end{tabular}}
\def\btbl{\begin{table}}
\def\etbl{\end{table}}
\def\bfig{\begin{figure}[htb]}
\def\efig{\end{figure}}
\def\bpic{\begin{picture}}
\def\epic{\end{picture}}

\def\ga{\mathrel{\raise.3ex\hbox{$>$\kern-.75em\lower1ex\hbox{$\sim$}}}}
\def\la{\mathrel{\raise.3ex\hbox{$<$\kern-.75em\lower1ex\hbox{$\sim$}}}}
\def\gappeq{\mathrel{\rlap {\raise.5ex\hbox{$>$}}
{\lower.5ex\hbox{$\sim$}}}}
\def\lappeq{\mathrel{\rlap{\raise.5ex\hbox{$<$}}
{\lower.5ex\hbox{$\sim$}}}}

\def\gyr{{\rm \, G\kern-0.125em yr}}
\def\mev{{\rm \, Me\kern-0.125em V}}
\def\gev{{\rm \, Ge\kern-0.125em V}}
\def\tev{{\rm \, Te\kern-0.125em V}}

\renewcommand{\bar}{\overline}

\newcommand{\rst}[1]{\raise+.6ex\hbox{#1}}

\begin{document}

\begin{titlepage}

\setcounter{page}{1}

\vspace*{0.2in}

\begin{center}

\hspace*{-0.6cm}\parbox{17.5cm}{\Large \bf \begin{center}

Singlet Neighbors of the Higgs Boson

\end{center}}

\vspace*{0.5cm}
\normalsize

\vspace*{0.5cm}
\normalsize

{\bf Brian Batell$^{\,(a)}$, David McKeen$^{\,(b)}$, and Maxim Pospelov$^{\,(b,c)}$ }

\smallskip
\medskip
$^{\,(a)}${\it Enrico Fermi Institute and Department of Physics, University of Chicago, \\
Chicago, IL, 60637, USA}

$^{\,(b)}${\it Department of Physics and Astronomy, University of Victoria, \\
     Victoria, BC, V8P 5C2 Canada}

$^{\,(c)}${\it Perimeter Institute for Theoretical Physics, Waterloo,
ON, N2J 2W9, Canada}

\smallskip
\end{center}
\vskip0.2in
%\centerline{\large\bf Abstract}
\begin{abstract}
The newly discovered resonance at 125 GeV has properties consistent with the 
Standard Model (SM) Higgs particle, although some production and/or decay channels 
currently exhibit $O(1)$ deviations. 
We consider scenarios with a new scalar singlet field with couplings to electrically charged vector-like matter, focusing particularly on the case when the singlet mass lies within a narrow $\sim$ few GeV window around
the Higgs mass. Such a `singlet neighbor' presents novel mechanisms for modifying the observed properties of the Higgs boson. For instance, even a small amount of the Higgs-singlet mixing can lead to a significant enhancement of the apparent diphoton rate. Alternatively, the Higgs may decay into the nearby singlet, along with a very light, very soft mediator particle, in which case there can be $O(1)$ enhancement to the apparent diphoton rate even for $\sim$ TeV-scale charged vector-like matter.  We also explore models in which vector-like fermions mix with the SM leptons, causing the simultaneous enhancement of $\gamma\gamma$ and suppression of  $\tau\bar \tau$ Higgs branching ratios. 
Our scenario can be tested with the accumulating LHC data by probing for the di-resonance structure of the 125 GeV diphoton signal, as well as the relative shift in the resonance location between the diphoton and four-lepton modes.

\end{abstract}

\end{titlepage}

\section{Introduction}

Recent experimental developments at the LHC, combining 2011 and 2012 data sets, 
have firmly established the existence of a resonance at $\sim 125$ GeV~\cite{ATLAS,CMS}. 
At this point, the newly discovered boson has properties fully consistent with the elementary
Higgs boson of the Standard Model (SM). This conclusion has further independent support from the Tevatron~\cite{Tevatron}, 
where the excess of the observed data over background can also be interpreted as a signature of the Higgs boson 
with mass in the range from 115 to 135 GeV. Future improvements at the LHC will allow for the precision determination of 
many decay channels for the new particle (see, {\em e.g.},~\cite{Peskin}). With the Higgs mass tentatively determined to be 125 GeV, there 
are no free parameters left in the SM, and all decay chains of the Higgs boson can be unambiguously predicted~\cite{HHG,Djouadi}. Although at present the data are consistent with the ``normal" Higgs, future data may reveal 
serious deviations from the SM predictions.

Currently, among the most interesting trends seen at the LHC is the larger-than-expected 
diphoton rate at 125 GeV, as explored in~\cite{gammagrave,Dave}, and the lack of any evidence for the coupling of the new resonance to leptons,
manifested in lower-than-expected rates of tau pair production at the same invariant mass. 
If these deviations grow to a significant level, they would imply a 
``non-minimal" Higgs boson.  At this point, one approach would be to make a generic parametrization of 
the new boson couplings to the rest of the SM (as was done in numerous studies 
on the subject over the years~\cite{Oldstuff} and most recently 
in connection with the positive Higgs signal; see for example~\cite{RecentFits}), and perform fits to the coupling constants in 
light of the currently available data. An alternative route is to formulate 
UV-complete models that modify Higgs boson properties in a calculable way as a function of 
model parameters \cite{models}.  

In this paper, we explore  extensions of the Higgs sector by a singlet scalar field.  This singlet scalar may couple to new vector-like, electrically charged matter, which induces a sizeable effective coupling to photons.  Within this general setup, we discuss new mechanisms to enhance the effective $h\to \gamma \gamma$ rate.
We especially concentrate on the special case of a quasi-degenerate Higgs--singlet 
pair. Current data point to the new resonance around 125-126 GeV, but are not sufficient to determine whether 
this is a single resonance or a series of excitations. We show that a quasi-degeneracy in mass, $ \Delta M < $ few GeV, is still allowed by current data and furthermore provides new opportunities for enhancing the apparent diphoton rate.
We discuss two novel mechanisms utilizing such a nearby `singlet neighbor': 
\begin{itemize}
\item A singlet neighbor that is very weakly mixed with the Higgs will be produced in the same manner as the SM Higgs particle, though with a smaller rate. However, it can easily have an $O(1)$ branching ratio to photon pairs if it couples to light charged vector-like matter. The mixing angles required to enhance the apparent $h \rightarrow \gamma\gamma$ rate by a factor of $\sim 2$ through this mechanism are much smaller than those required in other scenarios with mixed Higgs-singlet and significant mass separation 
between scalar resonances.  
\item The Higgs may be connected to a nearby singlet neighbor by a very light mediator, and have a small branching fraction into the singlet-mediator pair. Again, the singlet can have an $O(1)$ branching ratio to diphotons, thereby enhancing the apparent $h \rightarrow \gamma\gamma$ rate. Remarkably, the charged vector-like matter in this case can be very heavy, $\sim$ TeV scale, in contrast to other mechanisms relying on charged matter to enhance the diphoton rate. The light mediators are very soft in the decay and thus do not affect the reconstruction of the diphoton pair. 
\end{itemize}
Furthermore, we show that if the quantum numbers of vector-like fermions allow them to couple to leptons, 
one can have a reduction in the effective rate of the Higgs boson decays to tau leptons. Therefore, the same set of new particles could plausibly be responsible for an enhancement of $h\to \gamma \gamma$ and a suppression of $h\to \tau \bar\tau$. We note that models with singlet scalars have been a focus of various phenomenological investigations in 
recent years~\cite{scalars}. In particular,  the modifications of the Higgs production and decay patterns in models with scalar singlets coupled to new colored/charged matter were explored recently in Refs. \cite{impostors}. 

Our models have two new generic signatures. First, the mass degeneracy of the Higgs and the singlet could potentially be  resolved with more data in the highest resolution channel, $h\to \gamma \gamma$.
Second, the simplest version of the model features relatively light vector-like particles that decay back to the SM 
fermions and gauge bosons. We calculate the direct production cross section 
of these states and discuss possible final states of their decay. Although 
no direct search of vector-like charged matter has been performed to date, 
we adjust our calculations to the existing searches of other exotic charged states
(direct chargino production), and argue that the mass range 
relevant for the modification of diphoton and ditau 
effective rates can be realistically probed by the 
LHC in the very near future.  However, models with light mediators 
may not be subjected to direct constraints 
for a long time due to a possibility of having heavy vector-like states. 

The rest of the paper is organized as follows. In the next section we introduce the model of a quasi-degenerate 
mixed Higgs--singlet pair and vector-like matter and discuss its generic consequences. 
In Sec.~\ref{sec:light_med}  we supplement the model with a light mediator and demonstrate that the charged states can be made significantly heavier, while still having a substantial impact on the apparent diphoton rate. 
In Sec.~4 we discuss explicit models of vector-like matter that couples to leptons, exploring their potential to modify the ditau rate. We also examine in this section several other phenomenological aspects of the vector-like fermions, including their potential production/decay signatures at the LHC and their implications for precision electroweak measurements. 
Section~\ref{sec:exp} provides a detailed analysis of the possible multi-resonance structure of the 125 GeV signal under the hypothesis of a nearby singlet scalar. 
We reach our conclusions in Sec.~\ref{sec:conclusions}.

\section{Singlet neighbor next door to Higgs}
\label{sec:singlet}
Adding a singlet real scalar $S$ is the simplest extension of the 
SM. Here we consider the Lagrangian
\begin{eqnarray}
\label{start}
-{\cal L} & \supset &  -\mu_H^2 H^\dag H + \lambda_H (H^\dag H)^4 + \frac{1}{2} \hat m_S^2 \hat S^2 + A H^\dag H \hat S  - B \hat S 
+ \frac{\lambda_S}{4} \hat S^4+ {\cal L}_{SF},  \nonumber \\
\end{eqnarray}
where in particular we have included the super-renomalizable Higgs portal coupling to $S$.
We are free to choose the parameter $B = A v^2/2$, where $v$ is the SM Higgs vacuum expectation value (vev),  
so that $S$ does not obtain a vev. The additional term, ${\cal L}_{SF}$, in (\ref{start}) 
is the coupling of $S$ to the vector-like (VL) matter $F$ that we shall 
specify later. The only condition that we place on $F$ at this point is that its mass does not come from the SM Higgs mechanism,
{\em i.e.} $m_F$ remains finite in the limit of $v\to 0$.
Expanding about the vacuum, we obtain the following mass terms
\begin{equation}
-{\cal L}_2 = \frac{1}{2}(2\lambda_H v^2)\hat h^2 + \frac{1}{2}\hat m_S^2 \hat S^2 + A v \hat  h \hat S.
\end{equation}
The mixing is thus described by the mass matrix
\begin{equation}
{\cal M} = \left( 
\begin{array}{cc}
2 \lambda_H v^2 &  A v \\
A v  &  \hat m_S^2
\end{array}
\right),
\end{equation}
which can be diagonalized by a rotation,
\begin{equation}
\left( 
\begin{array}{c}
\hat h \\
\hat S
\end{array} 
\right)
= 
\left( 
\begin{array}{cc}
c_\theta &  s_\theta \\
-s_\theta &  c_\theta
\end{array}
\right)
\left( 
\begin{array}{c}
 h \\
 S
\end{array} 
\right), ~~~~~~~~~ \tan 2\theta = \frac{2 A v }{\hat m_S^2 -2\lambda_H v^2}.
\end{equation}
While the general case of Higgs phenomenology has been worked out before \cite{scalars,impostors}, 
we shall concentrate on the limit in which $A v \ll 2 \lambda_H v^2 \sim \hat m_S^2$.
In this regime, the mass eigenvalues are simply
\begin{equation}
m_S \simeq \hat m_S, ~~~~~~~ m_h \simeq \sqrt{2\lambda_H }v^2.
\end{equation}
If the mixing angle is small, it can be approximated by
\begin{equation}
\theta  \simeq \frac{A v}{\hat m_Y^2 - 2 \lambda_H v^2} = 
 \frac{A v}{ m_S^2 - m_h^2} \simeq   \frac{A v}{2 m_h \Delta M } ,
\label{Yangle}
\end{equation}
while for $|A|v\gg  |\hat m_Y^2 - 2 \lambda_H v^2|$ the mixing angle is maximal, $\theta = \pm \pi/4$.
After diagonalization, $h, S$ couples to the currents 
\begin{eqnarray}
{\cal L}  & \supset &  \hat h J_h + \hat S J_S \\
& = &  h (\cos\theta  J_h- \sin\theta J_S)    +   S  ( \sin\theta J_h + \cos\theta J_S).
\end{eqnarray}

\subsection{Production and decay of $S$ and modification of apparent Higgs signal}

Due to the interaction with the Higgs current, the $S$ boson 
will be produced in the same way as the SM Higgs with a rate smaller than the Higgs by the square of the mixing angle:
\begin{equation}
\sigma_{pp\rightarrow S} = (\sin\theta)^2 \sigma_{pp\rightarrow h}^{\rm SM}.
\end{equation}
We shall assume that the mass separation of $S$ and $h$ is small enough not 
to be resolved at current statistics, but large enough so that there is no interference 
at the level of the amplitudes, $\Delta M \gg \Gamma_h, \Gamma_S$. We address the question of the allowed size of the splitting $\Delta M$  in Section~\ref{sec:exp}.

In order to achieve large variations in the apparent diphoton rate, the production cross section of $S$
should be at least as large as the cross section for $pp\rightarrow h \rightarrow \gamma \gamma$, implying 
\begin{equation}
\theta^2 \gtrsim {\rm Br}^{\rm SM}_{h\rightarrow \gamma \gamma} = 0.0023 ~~ \Longrightarrow ~~
\theta \gtrsim 0.05,
\label{cond1}
\end{equation}
where we take Higgs branching ratio at $m_h=125$ GeV. 
Using the expression for the mixing angle in Eq.~(\ref{Yangle}), we can estimate the required value of the 
trilinear $A$ parameter:
\begin{equation}
|A| \gtrsim  0.05~{\rm GeV} \left|\frac{\theta }{0.05}\right|  \left( \frac{m_h }{125~\rm GeV}\right)
 \left| \frac{\Delta M }{\rm GeV}\right|,
\end{equation}
while the sign of $A$ as well as $\Delta M$ can be arbitrary. 

The coupling to charged matter $F$ will create an extra decay channel for $S$. 
Since the non-SM charged particles would have to be heavier than $m_{S,h}/2$, loops of $F$ will open the $S\to \gamma\gamma$ channel, to which we assign 
the width $\Gamma^F_{S\to\gamma\gamma}$. 
The $S$ particle can decay  into a pair of photons, or it may decay into 
other SM final states through its coupling to the Higgs current. 

It is easy to write down the general formula for the modification of the apparent Higgs signal in a mass window 125 GeV$\pm \Delta M$, normalizing to the SM rates:
\begin{align}
\label{Rvalue}
R_{\rm f.s.} &= \frac{\sigma_{pp\rightarrow h}\times {\rm Br} _{h\rightarrow \rm f.s.}+
\sigma_{pp\rightarrow S}\times {\rm Br} _{S\rightarrow \rm f.s.}}
{\sigma_{pp\rightarrow h}^{\rm SM} \times {\rm Br} _{h\rightarrow \rm f.s.}^{\rm SM}} 
= \cos^2\theta \times \frac{ {\rm Br} _{h\rightarrow \rm f.s.}}{ {\rm Br} _{h\rightarrow \rm f.s.}^{\rm SM}}+ \sin^2\theta \times \frac{ {\rm Br} _{S\rightarrow \rm f.s.}}{ {\rm Br} _{h\rightarrow \rm f.s.}^{\rm SM}}
\\ \nonumber
&\simeq 1+\theta^2\frac{ {\rm Br} _{S\rightarrow \rm f.s.}}{ {\rm Br} _{h\rightarrow \rm f.s.}^{\rm SM}}
\end{align}
where ``f.s." stands for a generic final state, such as $\gamma\gamma$, $ZZ^*$, etc. The last step in the 
expression above assumes the smallness of the mixing angle and consequently neglects modifications of $h$ branching ratios due to the mixing with $\hat S$. 
The case when additional decay channels for 
$S$ are not present, and it decays back to the SM via mixing with the Higgs, corresponds to the 
trivial situation when ${\rm Br} _{S\rightarrow \rm f.s.}={\rm Br} _{h\rightarrow \rm f.s.}^{\rm SM}$, 
and $R_{\rm f.s.} =1$ for any final state. We shall address the question of 
what increase in statistics is required in order to resolve the two $h$ and $S$ 
states at given $\theta$ and $\Delta M$  in Sec.~\ref{sec:exp}. The relevance of two nearby scalars for the 125 GeV LHC signal 
was recently discussed in the context of the NMSSM in Ref. \cite{Gunion:2012gc}.

In the model we consider, the diphoton branching is very simple in the limit of small mixing angle,
\be
{\rm Br}_{S\to \gamma\gamma}\left(\sqrt{{\rm Br}_{h\to\gamma\gamma}^{\rm SM}}\la\theta\la 1\right) 
\simeq  \frac{\Gamma^F_{S\to\gamma\gamma}}{\theta^2 \Gamma^{\rm SM}_{h,~\rm tot.}},
\label{modbr}
\ee
where we have also assumed that $\Gamma_{S\to\gamma\gamma}^F\sim\Gamma_{h\to\gamma\gamma}^{\rm SM}\ll \Gamma^{\rm SM}_{h,~\rm tot.}$.  At a generic $O(1)$ mixing angle, the expression is more complicated due to the interference of $F$ and $h$ mediated 
$S$ decays to diphotons at the amplitude level. For this broad range of mixing angles, all 
expressions simplify to $\theta$-independent combinations:
\be
\label{Rgg}
R_{\gamma\gamma} \simeq 1 + \frac{\Gamma^F_{S\to \gamma\gamma}}{\Gamma^{\rm SM}_{h\to\gamma\gamma}};
~R_{\rm other~f.s.} \simeq 1.
\ee
Thus, if the $\Gamma^F_{S\to\gamma\gamma}/\Gamma^{\rm SM}_{h\to\gamma\gamma}$ is sizable, 
the apparent diphoton rate will be enhanced.

\subsection{Singlets coupled to electrically charged VL matter} 
\label{sSF}

For a 125 GeV Higgs, the partial decay width of the Higgs into a pair of photons is given by
$\Gamma^{\rm SM}_{h\rightarrow \gamma \gamma} \simeq 9.3\times 10^{-6}$~GeV. 
In order to generate a comparable contribution to the SM, we introduce electrically charged VL matter
coupled to $S$, 
\be
{\cal L}_{SF}  \supset y_{SF}S\bar F_L F_R + ({\rm h.c.}) ~~ {\rm or }~~ A_{SF} S F^*F.
\label{SF}
\ee
The first option is a VL fermion with Yukawa coupling $y_{SF}$, while the second option is a 
scalar with trilinear coupling $A_{SF}$. These new fermions or scalars can belong to various 
representations of $SU(2)_L \times U(1)_Y$, though we consider $SU(3)_c$ singlets in order to avoid additional strong production and decay channels.  

The partial decay width for $S\to\gamma\gamma$ mediated by the charged fermion or scalar loop is given by
\begin{equation}
\Gamma_{S\rightarrow \gamma \gamma}^{\rm fermion} = y_{SF}^2
\frac{N^2 Q^4 \alpha^2  m_S^3 }{256 \pi^3 m_F^2} | A_{1/2}(\tau_F) |^2 ; ~~
\Gamma_{S\rightarrow \gamma \gamma}^{\rm scalar} = A^2_{SF}
\frac{N^2 Q^4 \alpha^2  m_S^3 }{1024 \pi^3 m_F^4} | A_{0}(\tau_F) |^2,
\end{equation}
where $Q$ is the charge of $F$, $N$ is a possible multiplicity factor in case $F$ has additional quantum numbers
other than spin, $\tau_F = m_S^2/4 m_F^2$, and $A_{0,1/2}$ are the loop functions which are defined in 
the review \cite{Djouadi}. We have also assumed $CP$ conservation, so that $y_{SF}$ is explicitly real. 
In the limit of heavy charged particles, the loop functions take the values $A_{1/2} \rightarrow 4/3$
and $A_0\rightarrow 1/3$, and we may write the effective increase in the diphoton rate coming from $S$ as 
\begin{eqnarray}
R_{\gamma \gamma}-1\sim
0.3\times N^2 Q^4  \left( \frac{m_S}{125~{\rm GeV}}  \right)^3  \left\{
\begin{array}{c}
(y_{SF}/2)^2\, (150~{\rm GeV}/m_F)^2\\
(A_{SF}/1.25~{\rm TeV})^2\, (150~{\rm GeV}/m_F)^4
\end{array}
\right..
\label{Y2gam}
\end{eqnarray}

Clearly, a large modification of the apparent Higgs decay rate to diphotons
would require either Yukawa or trilinear couplings on the borderline of 
perturbativity, or large multiplicities or charges of the VL states (for a realization of 
the latter idea, see \cite{Cline}). Even with the absence of the strong production channels, the 
$F$ states in the mass range of a few hundred GeV should be copiously produced at the LHC.  We discuss constraints from this in Sec.~\ref{sec:VLmatter}. 

\section{Higgs connected to a singlet neighbor by a light mediator}
\label{sec:light_med}

We now modify our model, and introduce a light mediator $X$ that connects $h$ and $S$ states and remove the direct mixing term $ASH^\dagger H$. 
Consider adding two real gauge singlet scalars $S$, $X$,  with the following Lagrangian
\begin{eqnarray}
\nonumber
-{\cal L} & \supset & \frac{1}{2} \hat m_X^2  \hat  X^2 +  \frac{1}{2} \hat m_S^2 \hat S^2 + \lambda H^\dag H \hat  X \hat S + {\cal L}_{XSF} \\
& = & \frac{1}{2}\hat m_X^2  \hat X^2 +  \frac{1}{2}\hat m_S^2 \hat S^2 +\frac{1}{2} \lambda v^2 \hat X \hat S + 
 \lambda v h \hat  X \hat S  + \frac{1}{2} \lambda h^2 \hat X \hat  S  + {\cal L}_{XSF}, 
\label{XSFmodel}
\end{eqnarray}
where again ${\cal L}_{XSF}$ stands for singlet interactions with charged VL matter.
The mass mixing of the two singlets is described by the matrix
\begin{equation}
{\cal M} = \left( 
\begin{array}{cc}
\hat m_X^2 &  \lambda v^2/2 \\
\lambda v^2 /2 &  \hat m_S^2
\end{array}
\right).
\end{equation}
We diagonalize the system by a rotation 
\begin{equation}
\left( 
\begin{array}{c}
\hat X \\
\hat S
\end{array} 
\right)
= 
\left( 
\begin{array}{cc}
c_\theta &  s_\theta \\
-s_\theta &  c_\theta
\end{array}
\right)
\left( 
\begin{array}{c}
 X \\
 S
\end{array} 
\right), ~~~~~~~~~ \tan 2\theta = \frac{\lambda v^2}{m_S^2 - m_X^2}.
\end{equation}
We will always be working in the limit of $\hat m_X^2 \ll \lambda v^2 /2 \ll \hat m_S^2$. In this limit, the mixing angle is small, and is approximated by
\begin{equation}
\theta \simeq \frac{\lambda v^2}{2 \hat m_S^2}. \label{theta}
\end{equation}
The mass eigenvalues are approximately
\begin{equation}
\label{masseig}
m_X^2  = \hat m_X^2 - \frac{\lambda^2 v^4}{ 4 \hat m_S^2}, ~~~~~ m_S^2 \simeq \hat m_S^2. 
\end{equation} 
There will be some tuning coming from the need to have a mass eigenvalue   $m_X < 1$ GeV, as we will discuss below.

After diagonalizing the system, the interaction Lagrangian is given by
\begin{equation}
{\cal L} \supset
 \lambda v h   X  S -  \theta \lambda v  h X^2 + \theta \lambda v  h S^2  + {\cal L}_{XSF},
\end{equation}
where we have omitted the quartic scalar interactions. We now discuss several consequences of the model. We have in mind a spectrum of $m_h \sim 126$ GeV, $m_S\sim 125$ GeV, and $m_X < 1$ GeV.

\subsection{ New Higgs decay channel,  $h \rightarrow XS$}

Given the light mediator $X$, the apparent Higgs decay properties can be modified not by mixing but by the decay $h \rightarrow XS$.  This new channel can be very important for the apparent rate of $h\rightarrow \gamma \gamma$ if 
$\Gamma_{h\to XS} > \Gamma_{h \rightarrow \gamma \gamma}$ because it is easy to arrange that, once produced, the $S$ state is kinematically 
forced to decay to photons.  For this to mimic the diphoton signal, $X$ should be very light, $m_X  \lesssim $ GeV, in which case it will be very soft and not affect reconstruction of the diphoton pair. 
For a related discussion of an increase in the diphoton rate from the Higgs boson decaying to singlets that then decay to two photons, see~\cite{Dave,Dobrescu:2000jt}.

The $R$ ratio, in case when two states $S$ and $h$ cannot be resolved, is given by 
\be
R_{\gamma\gamma} = 1 + \frac{{\rm Br}_{h\to SX}}{ {\rm Br}^{SM}_{h\to \gamma \gamma}}
\times {\rm Br}_{S\to \gamma \gamma},
\ee
where we also assume that $\Gamma\left(h\to SX\right)$ does not exceed the SM Higgs width.

The partial decay width for  $h \rightarrow SX$ in the relevant kinematic regime
 $m_X \ll \Delta M \equiv m_h - m_S \ll m_h$ is given by
\begin{eqnarray}
\Gamma_{h\rightarrow XS} &  \simeq  & 
\frac{\lambda^2 v^2 \Delta M}{8 \pi m_h^2} \nonumber \\
& = & 10^{-5}~{\rm GeV} \left( \frac{\lambda }{0.01}\right)^2 \left( \frac{\Delta M}{ \rm GeV}\right) \left( \frac{126~{\rm GeV}}{m_h}\right)^2.
\end{eqnarray}
We see that in order to obtain a comparable new contribution to the standard 
$h\rightarrow \gamma \gamma$ rate, we must require having a coupling of $\lambda \sim 0.01$ or larger.
A coupling $\lambda \sim 0.01$ implies a mixing angle (\ref{theta}) of size
\begin{equation}
\theta \simeq 0.02 \left(\frac{\lambda  }{0.01} \right)\left( \frac{125~{\rm GeV}}{m_Y }  \right).
\end{equation}

\subsection{The degree of tuning in the singlet sector}
A coupling $\lambda \sim 0.01$ is large enough so that electroweak symmetry breaking makes an important contribution to the masses of $X$, $S$. There is a contribution to the physical $X$ mass given in Eq.~(\ref{masseig}) of
\begin{equation}
\label{tune}
\frac{\lambda^2 v^2 }{4 \hat  m_S^2} \sim O(5 ~ \rm GeV^2).
\end{equation}
To obtain a physical mass of $m_X = 0.5$ GeV, we must tune the bare mass parameter $\hat m_X^2$ so that there is a cancellation in  Eq.~(\ref{masseig}) at the $\sim 5  \%$ level. 

The tree-level tuning can be reduced if $\lambda H^\dagger H\hat X\hat S$ is ``traded" for a higher-dimensional effective operator $\frac{1}{\Lambda^2}\hat X\hat S\Box H^\dagger H$, that gives the same
effective $XSh$ vertex if $m_h^2/\Lambda^2=\lambda$, which implies the need in UV completion 
at a TeV scale. Although direct mass mixing between $\hat X$ and $\hat S$ is 
now absent, the tuning in this model can 
reappear as the loop-induced correction to the mass of $X$ scalar. 

In a modification of model (\ref{XSFmodel}),  $X$ and $S$ can be 
considered as real and imaginary part of a complex scalar field $\Phi$, 
$\Phi = 2^{-1/2}S\exp(iX/\langle S\rangle)$, charged 
under the global PQ-type symmetry. This  can lead to the masslessness of the $X$ scalar
in the limit of exact PQ symmetry.
However, in order to generate effective $XSh$ coupling, PQ-breaking terms must be 
introduced, $i\lambda( H^\dagger H - v^2/2)\Phi\Phi$ etc, which reintroduce the 
mass fine-tuning for $X$. 
A detailed realization of this scenario goes outside the scope of the present work.

\subsection{ Kinematic entrapment of $S$ and strong enhancement of  $S\rightarrow \gamma \gamma$}

If the VL states to which  $S$ is coupled are made very heavy, this
scalar can decay through an off-shell Higgs. Since the Higgs is just barely off-shell due to the small splitting, 
this decay mode can be small, but not negligible. 

The partial decay width for the dominant process of this kind, 
$S \rightarrow X h^* \rightarrow X \bar b b$, is given by
\begin{equation}
\Gamma_{S\rightarrow X \bar b b} = \frac{1}{\pi}\int_{4m_b^2}^{(m_S \! - \!m_X)^2} 
\!\!\!\!\!\!\! \!\!\! ds \frac{\sqrt{s}}{(s-m_h^2)^2 + m_h^2 
\Gamma_h^2} \Gamma_{S\rightarrow X h^*}(s) 
\Gamma_{h^*\rightarrow \bar b b}(s) ,
\end{equation}
where we have defined
\begin{eqnarray}
\Gamma_{S\rightarrow X h^*}(s) & = &
\frac{\lambda^2 v^2}{16 \pi m_S}
\lambda^{1/2}\left( 1, \frac{s}{m_S^2}, \frac{m_X^2}{m_S^2}\right), \\
\Gamma_{h^*\rightarrow \bar b b}(s) & = &  \frac{3 y_b \sqrt{s} }{16 \pi} \left( 1-\frac{4 m_b^2}{s}\right)^{3/2}.
\end{eqnarray}
In the limit $m_X, \Delta M, m_b \rightarrow 0$, we obtain the approximate expression
\begin{eqnarray}
\Gamma_{S\rightarrow X \bar b b} & \simeq &  \frac{3 y_b^2 \lambda^2 v^2}{256 \pi^3 m_h}\left[ \log\left( \frac{m_h}{ 2 \Delta M}\right)-2\right] \nonumber \\
& \approx &  10^{-8}~{\rm GeV} \times c\left( \frac{\lambda}{0.01}  \right)^2 \left( \frac{126~{\rm GeV}}{m_h } \right),
\end{eqnarray}
where $c$ is an $O(1)$ coefficient depending on $\Delta M$. We see that this decay width
can be up to three orders of magnitude below $\Gamma_{h\to \gamma\gamma}^{\rm SM}$.

Returning to our results for the $\gamma\gamma$ rate due to the couplings to fermions, we can see that 
${\rm Br}_{S\to \gamma\gamma}\simeq 1$ condition can be satisfied for $\lambda \simeq 0.01$ 
as long as 
\begin{eqnarray}
\Gamma_{S\rightarrow \gamma \gamma} \ga \Gamma_{S\rightarrow X \bar b b} ~~\Longrightarrow~~
 \left(\frac{y_{SF}}{2}\right)^2N^2 Q^4 
\left( \frac{3~{\rm TeV}}{ m_F } \right)^2  \ga 1.
\label{secluded}
\end{eqnarray}
Needless to say, this is far less extreme choice of parameters than (\ref{Y2gam}), which illustrates the fact that 
in models involving a light singlet $X$, the enhancement of apparent Higgs diphoton rate can be mediated by TeV scale charged particles, well outside of the current LHC reach. 

\subsection{Phenomenology of the light $X$}

The phenomenology of light $X$ can depend quite significantly on whether it directly couples to VL matter $F$.
If direct coupling to $F$ is absent, then $X\to\gamma\gamma$ decay is 
mediated by the small mixing of $X$ and $S$. 
The width ratio of $X$ to $S$ decays into photons is given by
\begin{eqnarray}
\frac{\Gamma_{X\rightarrow \gamma \gamma}}{ \Gamma_{S\rightarrow \gamma \gamma} }
& = & \theta^2 \left(  \frac{m_X}{m_S}\right)^3 \nonumber 
\\
& \simeq  & 2.5 \times 10^{-11} \times \left( \frac{\theta }{0.02}\right)^2 
\left(\frac{m_X}{0.5~{\rm GeV} }\right)^3 \left(\frac{125~{\rm GeV} }{m_S}\right)^3 .
\end{eqnarray}
If this is the dominant decay mode, then lifetime is given by 
\begin{eqnarray}
c\tau_X & = & \frac{144\pi^3 m_F^2}{\theta^2 N^2 Q^4 \alpha^2 y_{SF}^2 m_X^3  } \nonumber \\
& = & 20~{\rm m}\times \frac{1}{N^{2} Q^{4} y_{SF}^2}
\left( \frac{0.02}{\theta }\right)^2 
\left(\frac{0.5~{\rm GeV} }{m_X}\right)^3 \left( \frac{m_F}{250~{\rm GeV} }\right)^2 ,
\end{eqnarray}
so that the $X$ is naturally long lived, on the scale of the LHC experiments and larger. 

A significant increase of $X$ decay rate to photons can occur in models with PQ-like symmetry in the $F$-sector,
${\cal L}_{XSF} \supset M_F\bar F_L F_R \exp(iX/\langle S\rangle)$. Taking $\langle S\rangle \sim v$, one can estimate 
that $\Gamma_{X\rightarrow \gamma \gamma}/\Gamma_{S\rightarrow \gamma \gamma} 
\sim (  m_X/m_S)^3$, and the decay length for a semi-relativistic $X$ shrinks to a cm.  However, this does not have any impact on Higgs searches at the LHC since the photons produced in the decay of $X$ are extremely soft when $S$ and $h$ are nearly degenerate.

A model with relatively prompt decay of $X$ to photons may lead to additional signatures in $B$-physics. 
In particular, the Lagrangian (\ref{XSFmodel}) can be  extended to include the $\lambda_XXX H^\dagger H$ terms. 
Then Higgs-penguin--type diagrams will lead to a $B^+\to K^+XX\to K^+\gamma\gamma\gamma\gamma$ rare decay
mode. The rate has been calculated in Ref. \cite{BtoDM},
\be
{\rm Br}_{B^+\to K^+\gamma\gamma\gamma\gamma} \sim 2\times 10^{-4} \lambda_X^2.
\ee
The existing measurement of such a decay mode mediated by $2\pi^0$ \cite{BABARpaperthatnobodycites}
implies that  sensitivity at the level $O(10^{-6})$ can be obtained, which would in turn probe $\lambda_X^2 \ga 0.01$. 
The decays of $B$ with subsequent decays of $X$ outside the detector would probe $\lambda_X^2 \ga 0.05$. 
We note, however, that the scaling for $\lambda_X$ dictated by technical naturalness is $\lambda_X \sim \lambda^2$,
and therefore this coupling can be much too small to allow any chance for its detection in the $B$ system.

\section{ Vector-like matter  and modification of $h\to \tau \tau$}
\label{sec:VLmatter}

Our mechanisms of enhancing the diphoton rate with nearby singlet scalar particles relies on the presence of new VL matter $F$ that carries electric charge. As such, $F$ must arise from one or more multiplets charged under $SU(2)_L$ and/or $U(1)_Y$. For simplicity, we do not consider VL matter charged under the $SU(3)_c$. In this section we discuss the possible implications of such VL sectors charged under the electroweak group, and in particular point out that such VL states may cause a modification to the $h\rightarrow \tau  \tau $ signal at the LHC.  

The VL matter will generically be pair produced at colliders through electroweak processes. The signatures of the VL pairs, however, are more model dependent, and will be dictated by the coupling of $F$ to the SM matter fields. A minimal choice would be to restrict to VL quantum numbers such that  renormalizable couplings of $F$ to SM matter fields are present. A catalog of these quantum numbers was studied, for example, in Ref.~\cite{minimalmatter}. Even with this simplifying assumption, there is a plethora of possible signatures depending on the spin and electroweak quantum numbers, as well as the flavor of SM leptons and/or quarks present in the interaction with $F$. 

Rather than an exhaustive survey of the possible quantum numbers and signatures of VL matter,  we wish to speculate whether the VL matter itself may have interesting implications for Higgs physics. For example,  if these VL states are fermionic then it is plausible that they couple to  SM leptons $\ell$ at the renormalizable level via couplings of the form $F\ell$ or $F H \ell$. This causes mass mixing that ultimately leads to shifts in the couplings of the Higgs boson to leptons. Clearly the most interesting case to consider experimentally is when $F$ mixes with the $\tau$, as it is the heaviest lepton and thus has the largest leptonic coupling to the Higgs. It is also intriguing that the current LHC data shows a slight deficit in the $h\rightarrow \tau \tau$ channel. As we will see, there are viable models of VL matter that can cause such a deficit. The apparent $\tau$-lepton signal relative to the SM can be given by 
\be
\label{Rtt}
R_{\tau\tau} \simeq \frac{{\rm Br}_{h\to \tau \tau}}{{\rm Br}_{h\to\tau\tau}^{\rm SM}} + 
\theta^2 \frac{{\rm Br}^F_{S \to \tau \tau}}{{\rm Br}^{\rm SM}_{h\to \tau\tau}},
\ee
where we assume that the mixing angle is small and that the $h\tau\tau$ vertex 
can be modified by the presence of VL matter. Obtaining $R_{\tau\tau} <1$ 
would obviously require that both terms in (\ref{Rtt}) are less than one. In the remainder of this section we will evaluate 
 whether the reduction of the apparent $\tau$ rate is possible within a class of models containing VL fermions that mix with $\tau$. We will furthermore investigate the implications of these models for colliders and precision electroweak  measurements. For a recent study exploring the potential of VL fermions to enhance the diphoton rate, see Ref.~\cite{RecentVL}.

\subsection{Precision electroweak constraints on $F-\tau$ mixing}
\label{subsec:EWPD}

The VL leptons we will consider are subject to precision electroweak constraints. In particular such leptons can give contributions to the oblique parameters $S$ and $T$~\cite{oblique}, and perhaps more importantly, mixing between $\tau$ and VL leptons will lead to a non-universal shift in the 
$Z \bar \tau \tau $, and $W\tau\nu$ electroweak vertices. We now give a summary of these constraints. 

The $Z \bar \tau \tau $ vertex can be written as 
\begin{equation}
{\cal L} \supset \frac{g}{2c_W} Z_\mu \bar \tau \left(   g^\tau_V - g^\tau_A  \gamma^5 \right) \tau,
\end{equation}
where the vector and axial couplings are defined as $g_V^\tau = T_3-2 Q s_W^2$, $g_A^\tau = T_3$, with $s_W$ the sine of the weak mixing angle and $T_3 = -1/2$, $Q=-1$ for $\tau$. These tree level couplings are subject to radiative corrections which are encoded in the effective couplings  $\bar g_V^\tau =\sqrt{ \rho_\ell}(T_3 - 2 Q s_{{\rm eff},\ell}^2)$, $g_A^\tau = \sqrt{\rho_\ell}\,T_3$,  with $\rho_\ell = 1.005$, $s^2_{{\rm eff},\ell} =0.23128$~\cite{zpole}  assuming lepton-universality as in the SM. The SM predictions for the effective couplings are therefore $(\bar g_V^\tau)_{\rm SM} = -0.0370 \pm 0.0003$,~~$(\bar g_A^\tau)_{\rm SM} = -0.5012 \pm 0.00025$.  For non-universal scenarios such as the one we are considering, precision measurements on the $Z$-pole puts  constraints on these effective couplings~\cite{zpole}:
$\bar g_V^\tau = -0.0366 \pm 0.001,~~~\bar g_A^\tau = -0.5024 \pm 0.00064$.
These observables are weakly correlated, with a correlation coefficient of $\rho_{AV}^\tau = -0.07$.
In particular, we note that for the axial coupling $ g_A^\tau$, the SM prediction is greater than the experimental determination by more than 1$\sigma$, a fact which will become important when considering specific representations of VL leptons below.

There will also be a non-universal shift in the coupling $W\tau\nu$ vertex. Large deviations in this coupling are  constrained by tests of lepton-universality in lepton decays. From Ref.~\cite{hfag} we can obtain the bound on the ratio
$(g_\tau/g_\ell) =  1.0010 \pm0.0014$. In our models, the $W\tau\nu$ coupling will always be suppressed, and assuming there is no shift in the couplings of the $W$ to light leptons, the bound suggests we can tolerate a coupling as low as $g_{W\tau\nu} = 0.997$ at the $3\sigma$ level. In fact in the models we consider the shifts will always be much less than this.

Besides the non-universal vertex corrections, VL fermions will give contributions to the oblique parameters $S$ and $T$~\cite{oblique}. We use the updated Particle Data Group values~\cite{PDG} for a 125 GeV Higgs boson, $S=0.04\pm0.09$, $T=0.07\pm0.08$ with a correlation coefficient $\rho = 0.88$.

\subsection{Mechanisms for modifying $h\rightarrow \tau \tau$}
\label{subsec:mechanisms}

The models we consider have two built-in mechanisms for suppressing the rate of Higgs-to-tau leptons:
\begin{enumerate}
\item $\tau-F$ mixing: In this case significant mixing between VL fermions and $\tau$ leads to large modifications to the $h\tau\tau$ vertex. There are potentially large non-universal corrections to the $Z\tau\tau$ vertex that constrain these models.
\item $h-S$ mixing: In this case, the VL fermions mix only mildly with $\tau$, so as to suppress corrections to the $Z\tau\tau$ vertex. At the same time, $S$ has large off-diagonal couplings to the VL fermions, and through $S-h$ mixing a cancellation occurs in the physical $h\tau\tau$  coupling. 
\end{enumerate}

 From the point of view of the low-energy effective theory, integrating out heavy fermion VL 
partners may be phrased as a series of higher-dimensional operators. Some examples of the first  mechanism are given by the expression
\be
\label{masscorr}
{\cal L}_{\tau}\supset -y_\tau \bar \tau_L\tau_R{H}\left(1 +2c_1 \frac{H^\dagger H}{\Lambda^2} +...\right).
\ee
It is easy to see that if we truncate the series at these two terms, the negative $c_1$ is required, 
\be
R_{\tau\tau} = \left(\frac{1+ 3c_1v^2/\Lambda^2}{1+c_1v^2/\Lambda^2}\right)^2,
\ee
and even a $\sim$10\% mass correction translates into a factor of 2 suppression of the effective Higgs-to-tau decay rate. 

We will now illustrate both mechanisms with specific models.

\subsection{ A single VL fermion }

Consider first the simplest model, which contains a single new VL fermion $E$ with the quantum numbers of the the right-handed leptons of the SM: $E\sim (1,1,-1)$.
While in principle the VL fermions can couple to any possible combinations of the SM leptons, for simplicity, we will assume the couplings to electrons and muons are negligible and that $E$ only mixes with $\tau$.  As we will demonstrate, modifications of $h \rightarrow \bar \tau \tau$ are possible in this model due to the $S-h$ mixing mechanism described above. The other mechanism of $\tau-E$ mixing would lead to a large renormalization of the $Z\tau \tau$ vertex in conflict with precision $Z$-pole data, as we will discuss below. 

The Lagrangian  of the model is
\begin{eqnarray}
\label{tauE}
-{\cal L} 
& \supset &  
(\bar \ell_L,  ~\bar E_L)
\left\{
\left(
\begin{array}{cc}
 y_1 H & y_2 H \\ 
  0      &       M_E
\end{array}
\right) + 
\hat S \left(
\begin{array}{cc}
 0 & 0 \\ 
  y_{SE1} &  y_{SE2}
\end{array}
\right) 
\right\}
\left( 
\begin{array}{c} 
\tau_R \\
E_R 
\end{array} \right)
 \\
 & = & 
 (\bar \tau_L, ~\bar E_L)
 \left\{ 
 \left(
\begin{array}{cc}
 Y_1 & Y_2\\
  0 &M_E
\end{array}
 \right)
  + \frac{\hat h}{v}  
 \left(
\begin{array}{cc} 
Y_1 & Y_2\\ 
 0 & 0
\end{array}
 \right) + \hat S \left(
\begin{array}{cc}
 0 & 0 \\ 
  y_{SE_1} &  y_{SE_2}
\end{array}
\right) \right\}
 \left( 
 \begin{array}{c} 
 \tau_R \\
 E_R 
 \end{array} 
 \right), \nonumber 
\end{eqnarray}
where $\ell_L^T = (  \tau_L ,  \nu_L)$. Note that we have used a field redefinition to remove a possible vector-like mass term $\bar E_L \tau_R$ (see {\em e.g.}~\cite{Choudhury}). In the second line of (\ref{tauE}) we have 
retained only terms relevant for the neutral Higgs phenomenology, and defined $Y_i \equiv y_i v/\sqrt{2}$. For the scalar sector, we assume that $\hat h$ and $\hat S$ mix as described in detail in Section \ref{sec:singlet}.
In the simplifying case in which all of the entries are real, the mass eigenstates are obtained through 
separate left and right rotations for the fermions,
\begin{equation}
\left( 
\begin{array}{c}
\tau_{L,R} \\
F_{L,R} 
\end{array}
\right)  = 
\left( 
\begin{array}{cc}
c_{L,R}& s_{L,R}\\
-s_{L,R} & c_{L,R}
\end{array}
\right)
\left( 
\begin{array}{c}
\tau_{L,R} \\
F_{L,R} 
\end{array}
\right),
\end{equation}
where $s_{L,R} = \sin \theta_{L,R}$ are the mixing angles in the left and right sectors. These rotations will shift the $Z\tau\tau$ vertex, as described in Section~\ref{subsec:EWPD}. It is straightforward to work out the shifts of the vector and axial couplings in terms of the mixing angles $\theta_{L,R}$ in this model:
\begin{equation}
\delta g_V^\tau  = \delta g_A^\tau = \frac{1}{2} s_L^2.
\label{Edg}
\end{equation}
Therefore, in order to avoid strong constraints from from precision measurements sensitive to the $Z\tau\tau$ couplings, we must demand that the mixing is very small in the left sector. Combined with the requirement of a small eigenvalue for the physical $\tau$ mass, we are forced into the regime $Y_1 \lesssim Y_2  \ll  M$. In this regime, mixing angles and physical masses are then given by the approximate expressions
\begin{eqnarray}
m_\tau \simeq Y_1, ~~~~& ~~~~   m_E \simeq M_E, \nonumber \\
\theta_L \simeq  \displaystyle{\frac{Y_2}{M_E}},~~~~  & ~~~~ \theta_R \simeq  \displaystyle{\frac{Y_1 Y_2}{M_E^2}}.
\label{EmixAp}
\end{eqnarray}

We now explore the modifications of the apparent $h\rightarrow \tau \tau$ rate in this model. 
After diagonalizing the $\tau-E$ and $h-S$ sector, one can derive the following approximate expression for the coupling of the physical Higgs scalar $h$ to $\tau$
\begin{eqnarray}
y_{h\tau\tau} \approx  \frac{m_{\tau}}{v}  - \theta_{hS} \,  \theta_L \, y_{SE_1},
\end{eqnarray}
where we have employed the approximate mixing angles given in Eq.~(\ref{EmixAp}).
Taking $\theta_L \sim 0.04$, which gives a small but tolerable shift of $8\times 10^{-4}$ in 
$\delta g_{V,A}^\tau$, and $\theta_{hS} \sim 0.15$, $y_{SF_1} \sim 1$ we see that we can completely turn off the coupling $y_{h\tau\tau}$.  

\begin{figure}[htb]
\begin{center}
\includegraphics[width=.48\textwidth]{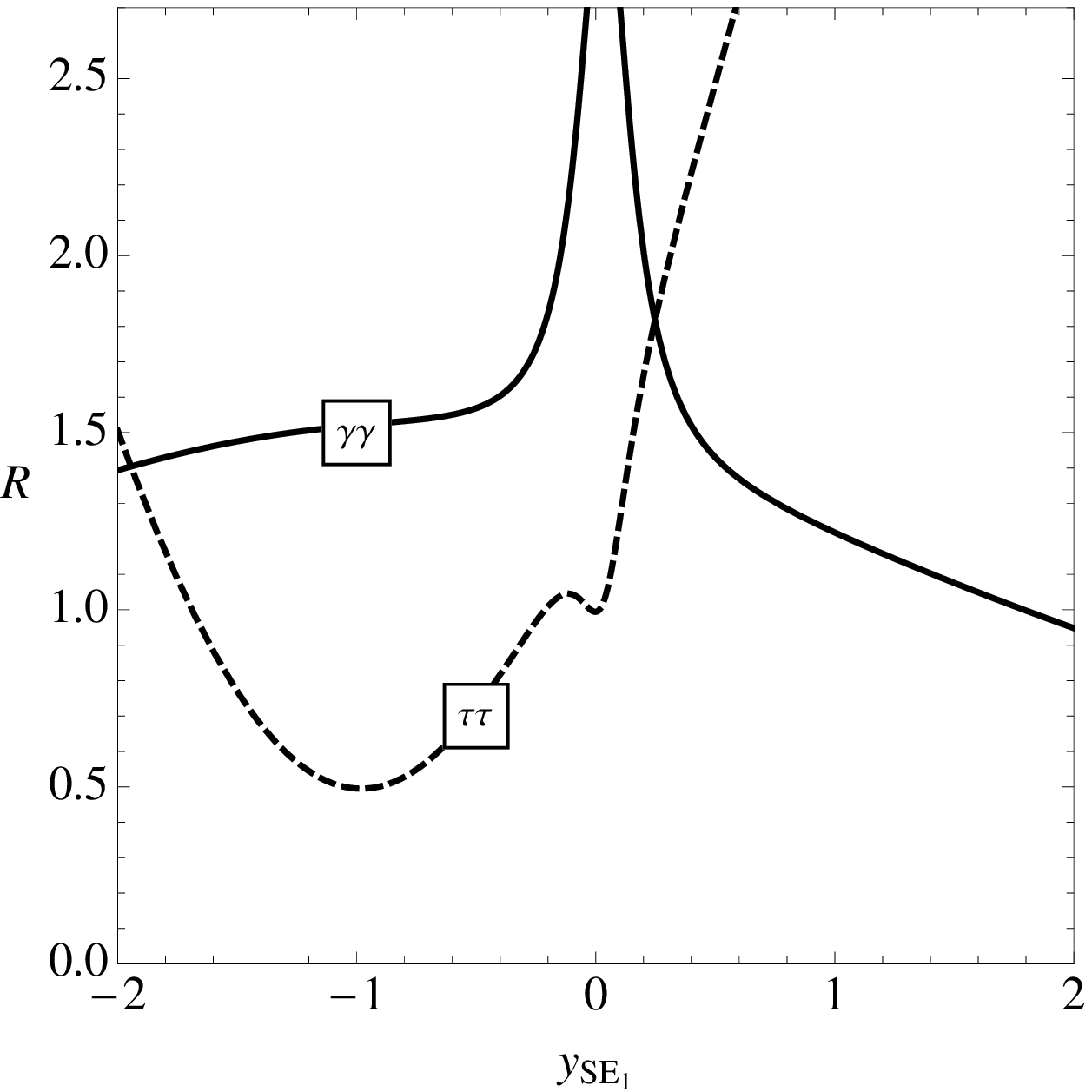} \quad
\includegraphics[width=.48\textwidth]{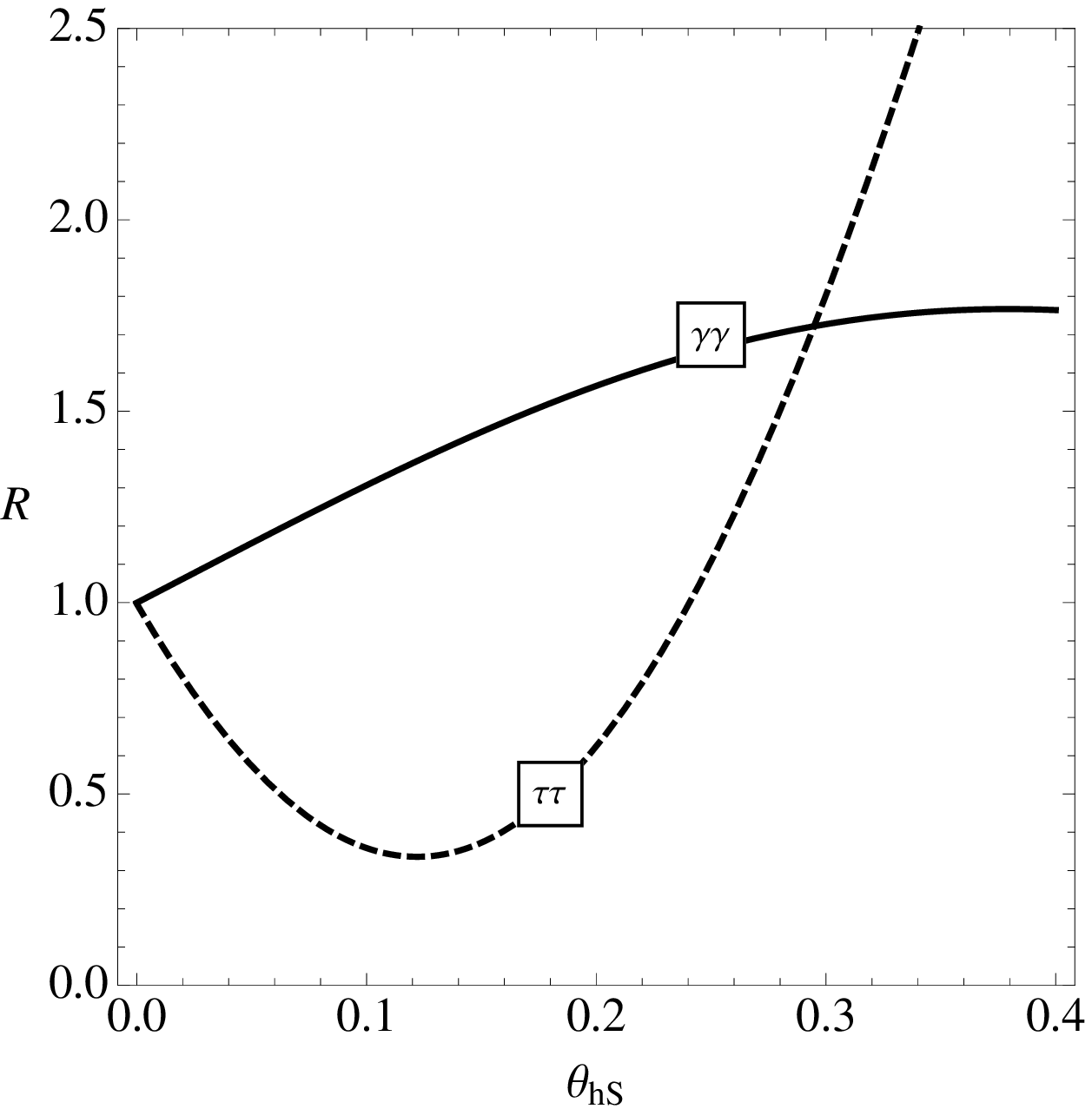} \quad
\caption{Apparent signal strength for the $\gamma \gamma$ channel (solid) and $\tau\tau$ channel (dashed) as a function of $y_{SE_1}$ (left) and $\theta_{hS}$ (right). We have fixed the remaining physical parameters to those indicated in the benchmark point in Table.~{\ref{Emodel}}.}
\label{fig:Etau}
\end{center}
\end{figure}

Even though we can suppress the coupling of the physical Higgs scalar to $\tau$, there will still be a nonzero apparent $h\rightarrow \tau\tau$ signal, since the singlet scalar $S$ also couples to $\tau$. Nevertheless the total apparent rate can be suppressed over a range of parameter choices. In Table~\ref{Emodel} we present a benchmark model which predicts a significant suppression of the signal in the $\tau\tau$ channel, as well as an enhancement in the $\gamma\gamma$ channel. This model is marginally consistent with $Z$-pole data in the tau sector.
We also show in Fig.~\ref{fig:Etau} the dependence of the $\gamma \gamma$ and $\tau\tau$ rates as a function of the Yukawa coupling $y_{SE_1}$ and the Higgs-singlet mixing angle $\theta_{hS}$.

\begin{table*}[t]
\begin{center}
\renewcommand{\arraystretch}{1.3}
 \begin{tabular}{ | c || c | c | c | c | c | c | }
\hline
\multicolumn{7}{|c|}{  $E \sim (1,1,-1)$  } \\ \hline \hline
$ {\rm Input} $ 
   & $Y_{1}$  &  $Y_{2}$  &  $M_E$  &   $y_{SE_1} $      &   $y_{SE_2}$   & $\theta_{hS} $   
  \\
 \hline \hline
& 1.7 &  5.2  & 130  & -1 & 3 & 0.18
\\  \hline \hline
$ {\rm Output} $ 
   & $m_\tau $  &  $m_E$   
 &   $R_{\gamma \gamma }$ &   $R_{\tau\tau}$   &   $\bar g_V^\tau  $  
  &   $\bar g_A^\tau $    
  \\     
     \hline
& 1.7   & 130  & 1.5 &   0.5 &  -0.0367   & -0.5004 \\
\hline
  \end{tabular}
  \label{Emodel}
\end{center}
\caption{  Benchmark model for single VL fermion  $E\sim (1,1,-1)$. The input parameters  $Y_{1}$, $Y_{2}$, $M_E$, and the output spectrum $m_\tau $,  $m_E$  have units of GeV, while
the remaining parameters/observables are dimensionless. }
\label{benchmark}
\end{table*}

It is worth mentioning that in this scenario, the enhancement in the $\gamma \gamma$ channel actually arises from the Higgs scalar $h$, which inherits a large coupling to the charged fermion $E$ due to its mixing with the $S$. Indeed, despite being produced with a sizable cross section, the $S$ particle in this setup dominantly decays to tau pairs. Thus, while we have taken $S$ to have a mass which is very close to that of $h$, there is no problem in separating $S$ from $h$ in terms of enhancing the diphoton signal. In this case, the $\tau\tau$ signal at $125$ GeV would completely disappear.

Due to the rigid correlation of mixing in the left sector in this model and nonuniversal corrections to the $Z\tau\tau$ couplings in Eq.~(\ref{Edg}), it is not possible to utilize large mixing between $\tau$ and $E$ to suppress the $h\rightarrow \tau \tau$ rate (the first mechanism described in Section~\ref{subsec:mechanisms} ). In fact, this occurs quite generally in models with a single VL fermion (though we have not done an exhaustive exploration of models). For example, in a model with a VL doublet fermion $L\sim (1,2,-1/2)$, the shift in the $Z\tau\tau$ vertex depends on size of the mixing angle in the right sector. As in the model just examined, requiring this mixing to be small ultimately implies that $\tau-L$ mixing cannot lead to a large modification to the  $h\tau\tau$ rate. Rather, the only way to obtain such a large modification is to rely on $h-S$ mixing. However, in models with more than one VL fermion, it is possible to simultaneously have small corrections to $Z\tau\tau$ couplings and large effects in $h\rightarrow \tau\tau$ through $\tau-F$ mixing, as we now explore in a model with two VL fermions.

\subsection{Two VL fermions}

While a single VL fermion cannot cause a modification to the Higgs-to-tau rate, it is possible if one considers two or more sets of VL fermions. 
To illustrate, we introduce a heavy VL lepton ``generation'', $L_{L,R} \sim (1,2,-3/2)$ and $E_{L,R} \sim (1,1,-1)$. As we will discuss below, the doublet with exotic hypercharge allows one to obtain better agreement with the $Z$-pole data related for the $\tau$ sector. 
The Lagrangian can be written in the following generic form, 
\be
\label{tauEL}
-{\cal L} \supset  (\bar \ell_L, ~\bar E_L, ~ \bar L_L)
\left\{ 
 \left(
\begin{array}{ccc} y_1 H& y_2 H &0\\m&M_E& y_3 \tilde H^\dag\\0 &  y_4 \tilde H & M_L 
\end{array}
 \right)
 + \hat S
  \left(
\begin{array}{ccc} 
0 &  0 & 0  \\  
y_{SE_1} & y_{SE_2}& 0 \\
0 &  0  & y_{SL_1} 
\end{array}
 \right)
  \right\}
\left( \begin{array}{c} \tau_R \\E_R \\L_R\end{array} \right),
\ee
where $m,~M_{E(L)}$ are bare VL 
mass parameters, and 
$y_i$ are Yukawa couplings. Again, we assume that $\hat h$ and $\hat S$ mix as described  in Section \ref{sec:singlet}.

This setup has clear advantages over a single VL fermion model, such as (\ref{tauE}), in terms of  its potential 
to modify the apparent Higgs decay rate to tau leptons through $\tau-L,E$ mixing: one can keep $m$ and $y_2 v$ small 
in comparison with $M_{E(L)}, y_{3(4)} v$, thus avoiding large renormalizations of the electroweak vertices 
of tau leptons. Furthermore, in this particular model one can obtain a small negative shift in $\delta g_A^\tau$ via mixing, moving the predicted value closer to the measured value inferred from $Z$-pole data, as discussed in Section~\ref{subsec:EWPD}. Finally, we note that there is an additional charge $Q=-2$ component $X$ which runs in the 
$h\gamma\gamma$ and $S\gamma\gamma$ loops, which can help enhance the apparent diphoton signal.

We show in Table~\ref{benchmark} a viable benchmark point which leads to a significant suppression of the $h\rightarrow \tau \tau$ branching ratio and an enhancement of the diphoton signal. 
The modifications of the apparent Higgs rates are as follows: 
\be
R_{\gamma\gamma} \simeq 1.4_h + 0.5_S \simeq 1.9;~~ R_{\tau\tau} \simeq 0.5_h + 0_S \simeq 0.5,
\label{ggtt}
\ee
where subscripts $h,S$ indicate the origin of the respective contributions. We see that in the example given above
the $\tau$ rate is lower than the that of the SM by a factor of 2, while gamma rate is increased by nearly a factor of 2. 
A sizable portion of the diphoton enhancement in this model is due to the mechanism discussed in Section~\ref{sec:singlet}, namely the nearly degenerate scalar with a small mixing to the Higgs portal and a coupling to the charged fermions in $L,E$.
\begin{table*}[t]
\begin{center}
\renewcommand{\arraystretch}{1.3}
 \begin{tabular}{ | c || c | c | c | c | c | c | c  | c| c| c| c| c|}
\hline
\multicolumn{12}{|c|}{ $L \sim (1,2,-3/2)$ , $E \sim (1,1,-1)$  } \\ \hline \hline
$ {\rm Input} $ 
   & $Y_{1}$  &  $Y_{2}$  &  $Y_{3}$ 
 &   $Y_{4}$ &   $m$      &   $M_E$     
 &   $M_L $      &   $y_{SE_1} $      &   $y_{SE_2}$      & $y_{SL_3}$ & $\theta_{hS} $   
  \\
 \hline \hline
& 3.1 &  7.5  & 130  & 80 & 35 & 250 &  250   & $ 0.4 $ & $2$ &  $2$ &  $0.05$  
\\  \hline \hline
$ {\rm Output} $ 
   & $m_\tau $  &  $m_E$  &  $m_L$ & $ m_X $
 &   $R_{\gamma \gamma }$ &   $R_{\tau\tau}$      &   $\Delta S$     
 &   $\Delta T $      &   $\bar g_V^\tau  $      &   $\bar g_A^\tau $    & $g_{W\tau\nu}$ 
  \\     
     \hline
& 1.7   & 148  & 357 &   250 &  1.9   & $ 0.5  $ & $ -0.04 $ &  $0.04$ &  $-0.0350$ 
& $-0.5016$ & $0.999$  \\
\hline
  \end{tabular}
\end{center}
\caption{  Benchmark model for VL lepton generation with $L\sim (1,2,-3/2)$, $E\sim(1,1,-1)$. The input parameters  $Y_{1}$, $Y_{2}$, $Y_{3}$, $Y_{4}$, $m$, $M_E$, $M_L $ and the output spectrum $m_\tau $,  $m_E$,  $m_L$, $m_X$  have units of GeV.
The remaining parameters/observables are dimensionless.  }
\label{benchmark}
\end{table*}

The model is in agreement with precision electroweak tests, with only modest corrections to the oblique parameters as indicated in Table~\ref{benchmark}. Regarding the $Z$-pole data related to the tau sector, the model gives a level of agreement comparable to that of the SM,  with $\bar g_{V,A}^\tau$ lying within 2$\sigma$ of the measured values.
In fact, the exotic doublet with hypercharge -3/2 has a slight advantage in this respect compared to a doublet with standard hypercharge -1/2. The latter model can only cause a positive shift in $\delta g_A^\tau$, which moves the prediction away from the measured value, while the exotic hypercharge model can cause either positive or negative shifts. We have investigated the model with standard hypercharge quantum numbers and have found that, for appropriate choices of parameters,  the model can be made marginally consistent with precision data while
still causing the desired effects in the apparent Higgs rates.

Concluding this section, it is fair to say that there are many other mechanisms that simultaneously affect
$R_{\gamma\gamma}$ and $R_{\tau\tau}$ that are not covered in this paper. 
In addition to the previously discussed case of VL fermions, one can achieve
similar effects by introducing additional Higgs doublets. One very plausible option is the 
two-Higgs doublet model with a lepton-specific doublet $H_l$. 
If the Higgs vev that gives masses to the $W$ and 
$Z$ mostly comes from the $H_q$ doublet, $\langle H_q\rangle \simeq v \gg \langle H_l\rangle $, then 
the observed 125 GeV state can be identified with the scalar $h_q$. Integrating out the $H_l$ fields 
may result in the structure (\ref{masscorr}) and thus a naturally suppressed Higgs
branching to tau pairs. To implement the increase in the diphoton rate due to the loop 
of charged Higgs states,  an unorthodox Higgs potential must be devised, 
so that the masses of the charged Higgs states from the $H_l$ sector have 
a ``flipped" dependence on the vev of $H_q$: $\partial m^2_\pm/\partial v <0$. 
The analysis of two-Higgs doublet models goes outside the scope of the present paper.

\subsubsection{Collider signatures of the VL matter sector }

Given the Lagrangian in Eq.~(\ref{tauE}), the $E$ fermions, transforming as $(1,1,-1)$ under the SM gauge group, will be pair produced at hadron colliders,
\be
\label{fs}
pp\to \gamma/Z^* \to E^+E^-. 
\ee
The $E^\pm$'s decay to $h\tau^\pm; ~ Z\tau^\pm;~ W^\pm \nu$ with rates in the ratio
\be
\label{fs}
\Gamma_{E^\pm\to h\tau^\pm} ~ \Gamma_{E^\pm\to Z\tau^\pm}:~ \Gamma_{E^\pm\to W^\pm \nu}\sim 1:1:2,
\ee
in the limit of large $m_E$.  In addition, if $S$ couples strongly to $\bar\tau E+{\rm h.c.}$, $E$ can decay via
\be
E^\pm\to S\tau^\pm\to\tau^+\tau^-\tau^\pm,~\gamma\gamma\tau^\pm.
\ee
This can lead to interesting signatures, such as $pp\to6\tau$, leading to many high energy, same-sign, same-flavor leptons, or $pp\to\tau^+\tau^-4\gamma$ with two pairs of photons having an invariant mass of 125~GeV.  In what follows, however, we concentrate on final states involving weak gauge bosons and leave the discussion of these novel signatures to future work.

\begin{figure}[htb]
\begin{center}
\includegraphics[width=.48\textwidth]{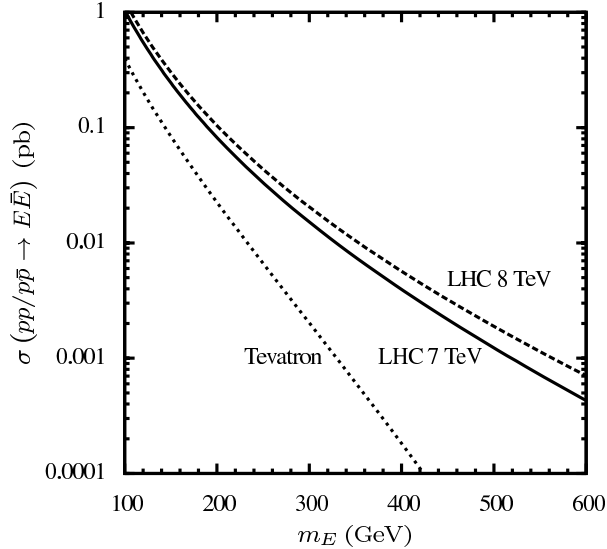}
\caption{The production cross section for $E^+E^-$ pairs at the LHC.  $E$ is a vector-like fermion with hypercharge -1.  For comparison we also show the cross section at the Tevatron.}
\label{fig:Eprod}
\end{center}
\end{figure}

The pair production cross section of charged $E$ fermions is given in Fig.~\ref{fig:Eprod}. For the model presented in Sec.~\ref{sSF}, where relatively light new charged particles are invoked to enhance the apparent Higgs diphoton rate, one should expect $O(10^3)$ charged pairs to have been produced at ATLAS and CMS. 

LEP sets a lower limit of the mass of changed particles of $\sim 105$~GeV.  For larger masses, we note that, while dedicated searches for exactly this type of charged particle have not yet been performed at the LHC, a recent ATLAS search for the production of charginos subsequently decaying to $W^\pm$ and a neutralino~\cite{AtlasConfNote} using data from the $\sqrt{s}=7$~TeV run is relevant.  The final state searched for in this analysis is $\ell^+\ell^{\prime -}$+missing energy where $\ell$ and $\ell^\prime$ each label a (possibly the same) light lepton $e$ or $\mu$ which can occur in $E^+E^-$ production when $E^\pm\to Z\tau^\pm,~W^\pm \nu$.  

\begin{figure}[htb]
\begin{center}
\includegraphics[width=.48\textwidth]{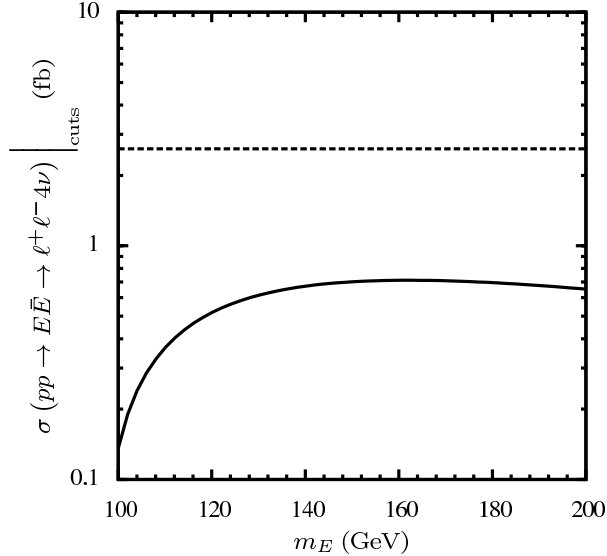}
\caption{The cross section for $pp\to E^+E^-\to\ell^+\ell^{\prime -}4\nu$ at the LHC with $\sqrt{s}=7$~TeV after making the cuts labelled SR-MT2 in~\cite{AtlasConfNote}.  The experimental upper limit of 2.6~fb is also shown.}
\label{fig:mt2}
\end{center}
\end{figure}

To estimate the reach of this analysis to the scenario described here, we generate $pp\to E^+E^-$ events with MadGraph 5~\cite{MadGraph} and decay the $E$'s to $W\nu$ and $Z\tau$.  We assume that ${\rm Br}_{E^\pm\to W^\pm \nu}=1/2$ and ${\rm Br}_{E^\pm\to Z \tau^\pm}=1/4$.  We then implement the cuts relevant to this final state (labelled SR-MT2) made in~\cite{AtlasConfNote}, ignoring any detector efficiencies.  The cross section after these cuts as a function of $m_E$ is presented in Fig.~\ref{fig:mt2}.  This should be compared with the upper limit found by ATLAS of 2.6~fb on non-SM cross sections in this region.

These rough estimates show that current data do not appear to rule out VL fermions light enough to have an appreciable effect on the Higgs diphoton rate as described in Sec.~\ref{sSF}, the LHC will be able to probe this scenario in the near future (similar conclusions were recently reached in Ref.~\cite{natural}).  Broadening the search to include additional final states including jets could allow for masses up to 250~GeV to be probed~\cite{4thgenlep}.  However, models like those described in Sec.~\ref{sec:light_med} are not constrained by such searches since they can accommodate large apparent increases to the Higgs diphoton rate with VL fermions whose masses are in the TeV range. 

\section{Searching for a singlet neighbor in the diphoton spectrum.}
\label{sec:exp}

If the apparent excess in the rate of $h\to\gamma\gamma$ observed at the LHC is due to the decay to photons of a nearly degenerate scalar, it is important to understand what size mass splitting is acceptable with current data and what will be able to be probed in the near future.

To roughly answer this question, we study the model described in Sec.~\ref{sec:light_med} and generate $pp\to h\to \gamma\gamma$ and $pp\to h\to SX$ events with $S\to\gamma\gamma$ using MadGraph 5~\cite{MadGraph}, fixing $m_h=126$~GeV and varying $m_S$.  The events are then processed by PYTHIA~\cite{Pythia} to add initial- and final-state showering and detector effects are approximated by PGS~\cite{PGS}.  We assume a large effective enhancement to the effective diphoton rate of $R_{\gamma\gamma}=2$, {\em i.e.} $\Gamma\left(h\to\gamma\gamma\right)=\Gamma\left(h\to SX\to\gamma\gamma X\right)$.  For simplicity, gluon fusion is the sole Higgs production mechanism we consider since it provides about 88\% of the Higgs production at the LHC at 8~TeV.

To understand the situation with current data, we generate 120 $h\to\gamma\gamma$ and $h\to SX\to\gamma\gamma X$ events (240 total) at each $m_S$ considered.  This corresponds to the number of events observed with an integrated luminosity of 5.9~fb$^{-1}$ at 8~TeV with a $\gamma\gamma$ reconstruction efficiency of $\sim 40\%$ with $R_{\gamma\gamma}=2$.  Additionally, we generate a further 6000 such events (12000 total) which corresponds to an integrated luminosity of 300~fb$^{-1}$ to estimate what effects increased data can have.

\begin{figure}[htb]
\begin{center}
\includegraphics[width=.48\textwidth]{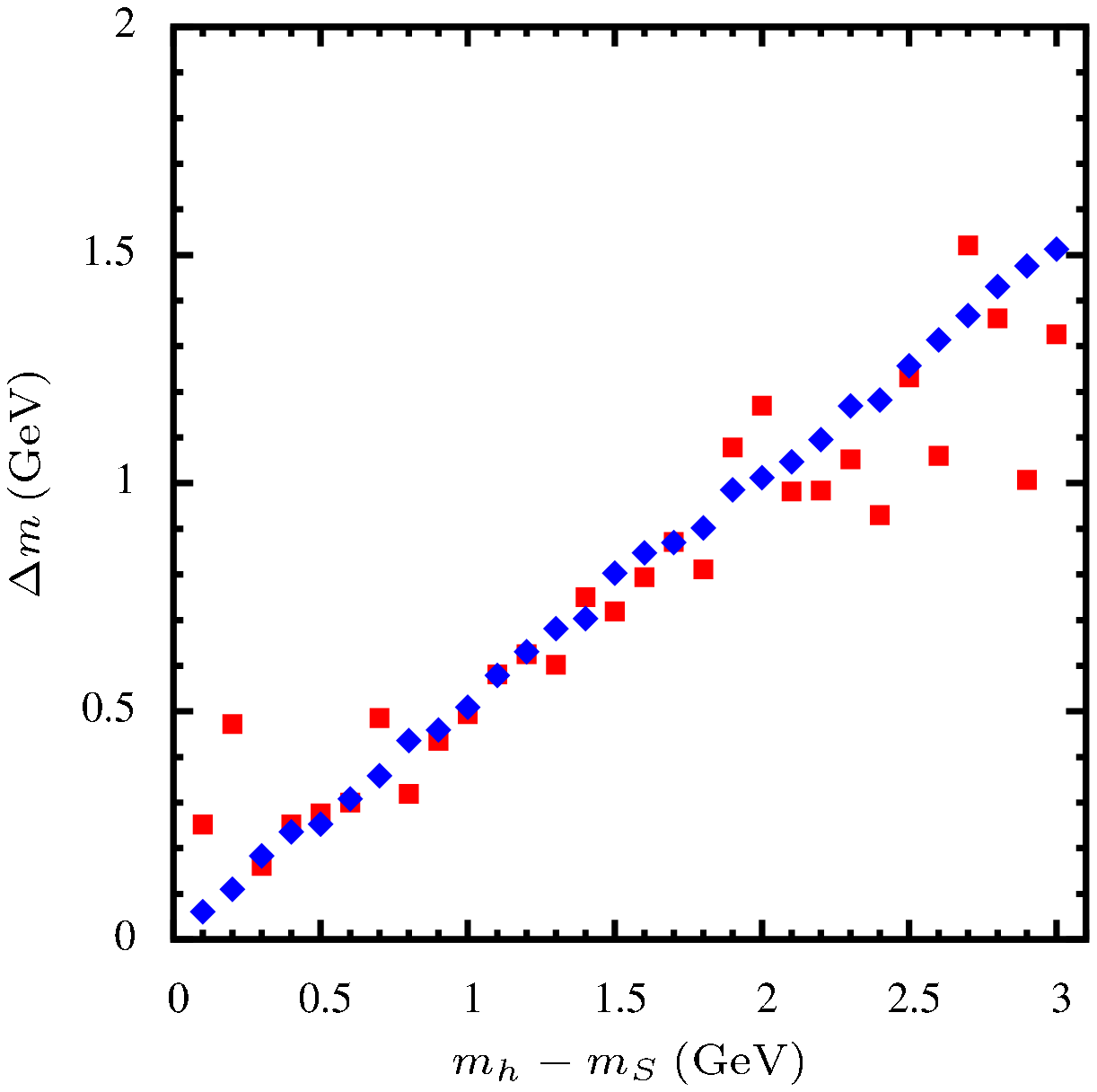}~~~~
\includegraphics[width=.48\textwidth]{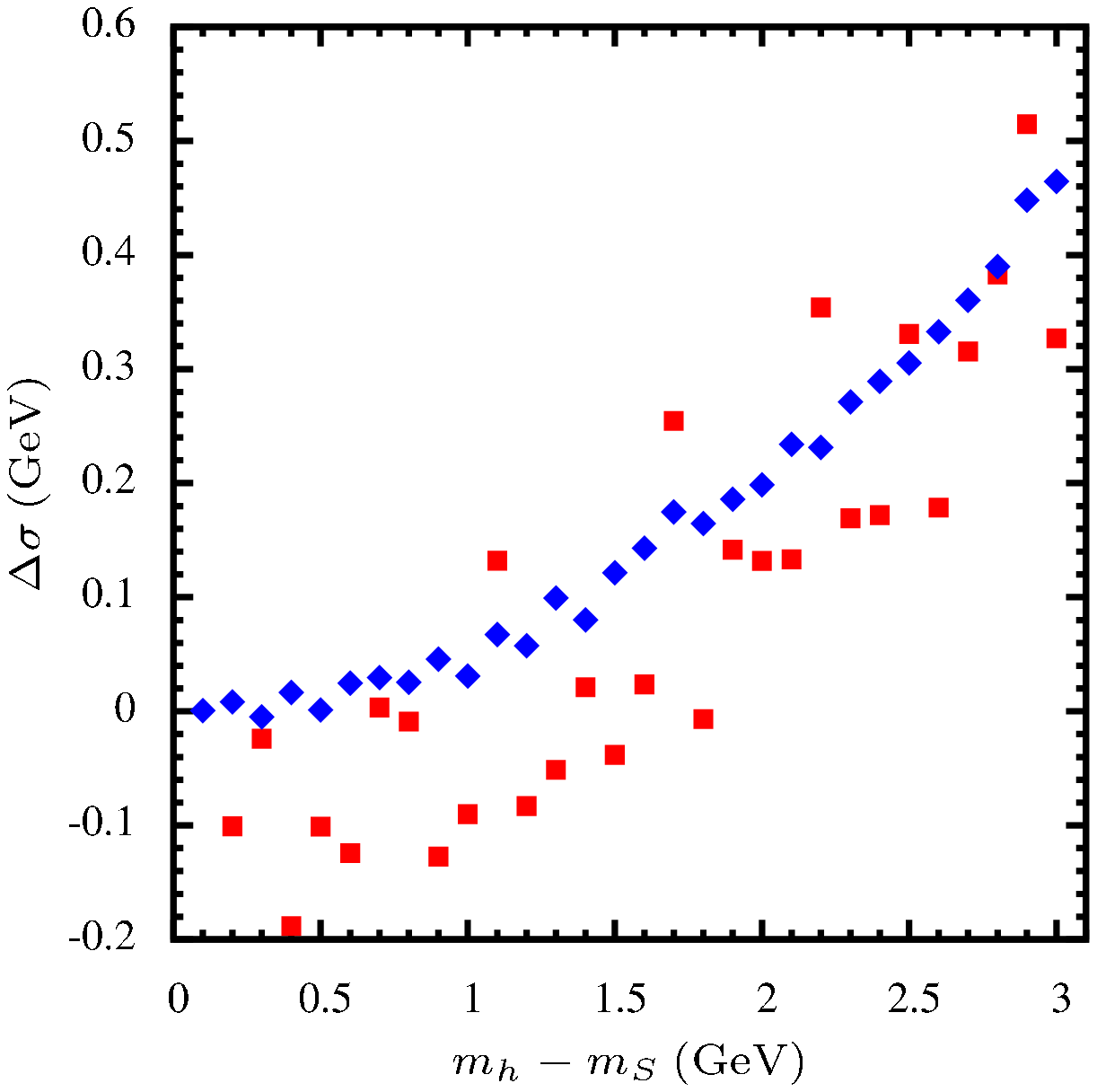}
\caption{The dependence of $\Delta m=m_h-m_{\rm fit}$ (left) and the  change in the width $\Delta\sigma$ (right) as a function of $m_h-m_S$  of the gaussian fitting the $M_{\gamma\gamma}$ signal from 120 $h\to\gamma\gamma$ and $h\to SX\to\gamma\gamma X$ events corresponding to 5.9~fb$^{-1}$ (red squares) and from 6000 $h\to\gamma\gamma$ and $h\to SX\to\gamma\gamma X$ events corresponding to 300~fb$^{-1}$ (blue diamonds). }
\label{fig:mfit}
\end{center}
\end{figure}

We present the results of fitting the diphoton invariant mass signals with a gaussian as a function of $m_h-m_S$. In the left panel of Fig.~\ref{fig:mfit} we see the shift in the fitted Higgs mass relative to the ÒtrueÓ Higgs mass as $m_h-m_S$ increases. Given the equal number of diphoton events coming from $h$ and $S$ the fitted mass simply tracks the average of $m_h$ and $m_S$. The change in the fitted width of the gaussian as a function of $m_h-m_S$ is shown in the right panel of Fig.~\ref{fig:mfit}.

\begin{figure}[htb]
\begin{center}
\includegraphics[width=.48\textwidth]{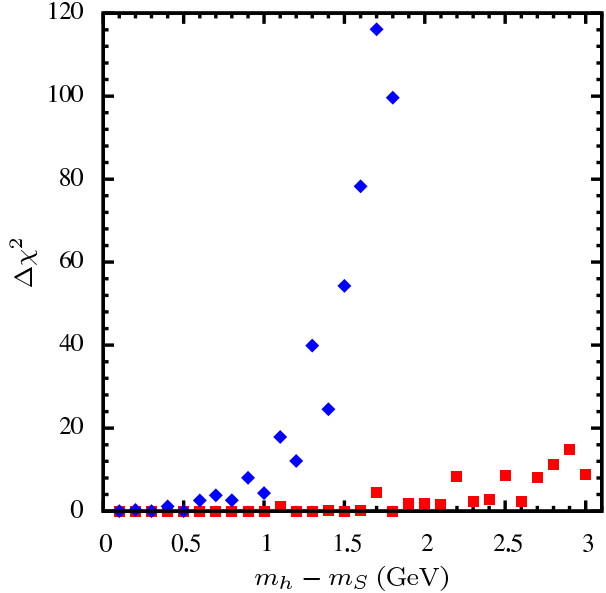}
\caption{The improvement in $\chi^2$ when fitting the $M_{\gamma\gamma}$ signal with two gaussians whose width is fixed to 2.1~GeV compared to a single gaussian whose width is fixed to 2.1~GeV with 120 $h\to\gamma\gamma$ and $h\to SX\to\gamma\gamma X$ events corresponding to 5.9~fb$^{-1}$ (red squares) and from 6000 $h\to\gamma\gamma$ and $h\to SX\to\gamma\gamma X$ events corresponding to 300~fb$^{-1}$ (blue diamonds).}
\label{fig:chi2}
\end{center}
\end{figure}

We also explore fitting the $M_{\gamma\gamma}$ signals with two gaussians whose width is fixed to 2.1~GeV (indicated by fits of the $M_{\gamma\gamma}$ shape from a single Higgs at 126~GeV in PGS).  The change in the $\chi^2$ when fitting the signal with two gaussians compared to fitting with a single gaussian whose width is fixed to 2.1~GeV for both 5.9~fb$^{-1}$ and 300~fb$^{-1}$ is shown in Fig.~\ref{fig:chi2}.

The presence of a nearly degenerate scalar could be deduced by observing a wider-than-expected $M_{\gamma\gamma}$ signal that is better fit by assuming two nearby resonances.  Additionally, the Higgs mass measured in the $\gamma\gamma$ channel could potentially disagree with that measured in the $ZZ^\ast$ channel.

These estimates indicate that the current diphoton excess seen at the LHC could potentially be caused by the two photon decays of a nearly degenerate scalar with a mass splitting $\Delta M=\left|m_h-m_S\right|$ as large as perhaps 2.5~GeV.  It appears that collecting 300~fb$^{-1}$ of data could probe $\Delta M\gsim 1$~GeV.

We should note that this analysis is very rough and definitive statements can only be made after a more full detector simulation is undertaken.  However, these estimates are given some amount of credence by observing that more complete studies indicate that the resolution on the Higgs mass in the diphoton channel is about 1~GeV~\cite{tdrs} and, in addition, the current experimental errors on the Higgs mass from this channel are a bit larger than this, roughly 2~GeV~\cite{ATLAS,CMS}.

\section{Conclusions}
\label{sec:conclusions}

We have explored the possible modifications to the apparent properties of the Higgs boson in a set of simple models in which the Higgs particle has a nearby singlet scalar `neighbor' that is coupled to charged VL fermions. 
Let us summarize the main consequences of the models we have considered: 

\begin{itemize} 

\item The presence of a second (or multiple) scalar state at $\sim$ 125 GeV is perfectly allowed by the current LHC data, and our analysis indicates that present sensitivity to $\Delta M$ does not exceed a few GeV. With the expected increase in statistics and the number of Higgs-like events collected at the LHC, it should be possible to resolve multiple states in the diphoton spectrum provided their splitting is greater than $\sim 1$ GeV.  

\item When the new state is populated via the mixing with the Higgs, even a relatively small 
mixing angle may lead to a significant increase in the apparent 
$h\to \gamma\gamma$ rate. This is true provided that model contains  
relatively light $O(100-300)$ GeV VL states, unless large charges/large 
multiplicities are introduced. We estimate that in the current dataset $O(10^3)$ of such VL states have been produced, but their detection depends rather sensitively on the decay modes, which are more model dependent. 

\item A light mediator particle connecting the Higgs and its singlet neighbor 
may lead to the additional channel of Higgs decay, $h\to XS$. 
An inefficient decay of $S$ via the off-shell Higgs enhances its branching ratios to other rare decay modes.
In particular, the apparent decay rate to photons can now be enhanced even with $O({\rm few ~TeV})$ scale charged particles,
which are currently outside of the LHC reach. (We note in passing 
that this model explicitly bypasses arguments of Ref. \cite{natural},
which argues that enhancement of $R_{\gamma\gamma}$ implies a low-scale of UV completion and relatively light 
VL fermions.)

\item In our models, the use of VL matter is essential for increasing the apparent rate of 125 GeV resonance 
to diphotons. We showed that  VL fermion that can couple to leptons can also lead to the reduction of the 
Higgs decay rate to tau leptons. 

\end{itemize}

\subsection*{Acknowledgements}

We would like to thank Drs. J. Albert and A. Ritz for helpful discussions. 
B.B. thanks the 2012 Santa Fe workshop {\it LHC Now}, sponsored by the 
Los Alamos National Laboratory, where part of this work was completed.
B.B. is supported by the NSF under grant PHY-0756966 and the DOE Early Career Award under grant
DE-SC0003930.
D.M. and M.P. are supported in part by NSERC, Canada, and research at the Perimeter Institute
is supported in part by the Government of Canada through NSERC and by the Province of Ontario through MEDT.


\begin{thebibliography}{99}

\bibitem{ATLAS} F. Gianotti, CERN Seminar,July, 4 2012. ATLAS-CONF-2012-093

\bibitem{CMS} J. Incandela, CERN Seminar,July, 4 2012.

\bibitem{Tevatron} S. Z. Shalhout, on behalf of the CDF  and D0, talk at the ICHEP2012, Melbourne, July 9, 2012. 

\bibitem{Peskin} 
  M.~E.~Peskin,
  %``Comparison of LHC and ILC Capabilities for Higgs Boson Coupling Measurements,''
  arXiv:1207.2516 [hep-ph].
  %%CITATION = ARXIV:1207.2516;%%

\bibitem{HHG} 
J. F. Gunion, S.  Dawson, H. E. Haber, G. L.  Kane,  {\em The Higgs hunter's guide},
Addison-Wesley, Reading, MA (1990).

\bibitem{Djouadi}
A.~Djouadi,
  %``The Anatomy of electro-weak symmetry breaking. I: The Higgs boson in the standard model,''
  Phys.\ Rept.\  {\bf 457}, 1 (2008)
  [hep-ph/0503172].

\bibitem{gammagrave}
M.~Carena, S.~Gori, N.~R.~Shah and C.~E.~M.~Wagner,
  %``A 125 GeV SM-like Higgs in the MSSM and the $\gamma \gamma$ rate,''
  JHEP {\bf 1203}, 014 (2012)
  [arXiv:1112.3336 [hep-ph]];
  %%CITATION = ARXIV:1112.3336;%%
  K.~Cheung and T.~-C.~Yuan,
  %``Could the excess seen at 124-126 GeV be due to the Randall-Sundrum Radion?,''
  Phys.\ Rev.\ Lett.\  {\bf 108}, 141602 (2012)
  [arXiv:1112.4146 [hep-ph]];
  %%CITATION = ARXIV:1112.4146;%%
  Z.~Kang, J.~Li and T.~Li,
  %``On Naturalness of the (N)MSSM,''
  arXiv:1201.5305 [hep-ph];
  %%CITATION = ARXIV:1201.5305;%%
  J.~J.~Heckman, P.~Kumar and B.~Wecht,
  %``The Higgs as a Probe of Supersymmetric Extra Sectors,''
  JHEP {\bf 1207}, 118 (2012)
  [arXiv:1204.3640 [hep-ph]];
  %%CITATION = ARXIV:1204.3640;%%
  A.~Azatov, R.~Contino, D.~Del Re, J.~Galloway, M.~Grassi and S.~Rahatlou,
  %``Determining Higgs couplings with a model-independent analysis of h ->gamma gamma,''
  JHEP {\bf 1206}, 134 (2012)
  [arXiv:1204.4817 [hep-ph]];
  %%CITATION = ARXIV:1204.4817;%%
  %\cite{Cohen:2012zz}
  T.~Cohen, D.~E.~Morrissey and A.~Pierce,
  %``Electroweak Baryogenesis and Higgs Signatures,''
  arXiv:1203.2924 [hep-ph];
  %%CITATION = ARXIV:1203.2924;%
  K.~Blum, R.~T.~D'Agnolo and J.~Fan,
  %``Natural SUSY Predicts: Higgs Couplings,''
  arXiv:1206.5303 [hep-ph];
  %%CITATION = ARXIV:1206.5303;%
 �A.~Arhrib, R.~Benbrik and C.~-H.~Chen, �%``H\to\gamma\gamma in the Complex Two Higgs Doublet Model,'' �arXiv:1205.5536 [hep-ph]; �%%CITATION = ARXIV:1205.5536;%%
  M.~Carena, S.~Gori, N.~R.~Shah, C.~E.~M.~Wagner and L.~-T.~Wang,
  %``Light Stau Phenomenology and the Higgs \gamma\gamma Rate,''
  arXiv:1205.5842 [hep-ph];
  %%CITATION = ARXIV:1205.5842;%%
  A.~G.~Akeroyd and S.~Moretti,
  %``Enhancement of H to gamma gamma from doubly charged scalars in the Higgs Triplet Model,''
  arXiv:1206.0535 [hep-ph];
  %%CITATION = ARXIV:1206.0535;%%
  M.~J.~Dolan, C.~Englert and M.~Spannowsky,
  %``Higgs self-coupling measurements at the LHC,''
  arXiv:1206.5001 [hep-ph];
  %\cite{Carena:2012xa}
  M.~Carena, I.~Low and C.~E.~M.~Wagner,
  %``Implications of a Modified Higgs to Diphoton Decay Width,''
  arXiv:1206.1082 [hep-ph];
  %%CITATION = ARXIV:1206.5001;%%
  J.~Chang, K.~Cheung, P.~-Y.~Tseng and T.~-C.~Yuan,
  %``Distinguishing Various Models of the 125 GeV Boson in Vector Boson Fusion,''
  arXiv:1206.5853 [hep-ph];
  %%CITATION = ARXIV:1206.5853;%%	
  %\cite{Chang:2012gn} 
  S.~Chang, C.~A.~Newby, N.~Raj and C.~Wanotayaroj,
  %``Revisiting Theories with Enhanced Higgs Couplings to Weak Gauge Bosons,''
  arXiv:1207.0493 [hep-ph];
  %%CITATION = ARXIV:1207.0493;%%
  %\cite{An:2012vp}
  H.~An, T.~Liu and L.~-T.~Wang,
  %``125 GeV Higgs Boson, Enhanced Di-photon Rate, and Gauged U(1)_PQ-Extended MSSM,''
  arXiv:1207.2473 [hep-ph];
  %%CITATION = ARXIV:1207.2473;%%
  %\cite{Bertolini:2012gu}
  %\cite{Craig:2012vn}
  N.~Craig and S.~Thomas,
  %``Exclusive Signals of an Extended Higgs Sector,''
  arXiv:1207.4835 [hep-ph].
  %%CITATION = ARXIV:1207.4835;%%
 

\bibitem{Dave} 
  P.~Draper and D.~McKeen,
  %``Diphotons from Tetraphotons in the Decay of a 125 GeV Higgs at the LHC,''
  Phys.\ Rev.\ D {\bf 85}, 115023 (2012)
  [arXiv:1204.1061 [hep-ph]].
  %%CITATION = ARXIV:1204.1061;%%
  
\bibitem{Oldstuff} C.~P.~Burgess, J.~Matias and M.~Pospelov,
  %``A Higgs or not a Higgs? What to do if you discover a new scalar particle,''
  Int.\ J.\ Mod.\ Phys.\ A {\bf 17}, 1841 (2002)
  [hep-ph/9912459]; 
A.~V.~Manohar and M.~B.~Wise,
  %``Modifications to the properties of the Higgs boson,''
  Phys.\ Lett.\ B {\bf 636}, 107 (2006)
  [hep-ph/0601212];
S.~Chang, R.~Dermisek, J.~F.~Gunion and N.~Weiner,
  %``Nonstandard Higgs Boson Decays,''
  Ann.\ Rev.\ Nucl.\ Part.\ Sci.\  {\bf 58}, 75 (2008)
  [arXiv:0801.4554 [hep-ph]]; I.~Low and J.~Lykken,
  %``Revealing the electroweak properties of a new scalar resonance,''
  JHEP {\bf 1010}, 053 (2010)
  [arXiv:1005.0872 [hep-ph]]. 


\bibitem{RecentFits} 
  C.~Englert, T.~Plehn, M.~Rauch, D.~Zerwas and P.~M.~Zerwas,
  %``LHC: Standard Higgs and Hidden Higgs,''
  Phys.\ Lett.\ B {\bf 707}, 512 (2012)
  [arXiv:1112.3007 [hep-ph]];
  %%CITATION = ARXIV:1112.3007;%%
  B.~Batell, S.~Gori and L.~-T.~Wang,
  %``Exploring the Higgs Portal with 10/fb at the LHC,''
  JHEP {\bf 1206}, 172 (2012)
  [arXiv:1112.5180 [hep-ph]];
  D.~Carmi, A.~Falkowski, E.~Kuflik and T.~Volansky,
  %``Interpreting LHC Higgs Results from Natural New Physics Perspective,''
  arXiv:1202.3144 [hep-ph];
  %%CITATION = ARXIV:1202.3144;%%
  M.~Klute, R.~Lafaye, T.~Plehn, M.~Rauch and D.~Zerwas,
  %``Measuring Higgs Couplings from LHC Data,''
  arXiv:1205.2699 [hep-ph];
  %%CITATION = ARXIV:1205.2699;%%
  J.~R.~Espinosa, M.~Muhlleitner, C.~Grojean and M.~Trott,
  %``Probing for Invisible Higgs Decays with Global Fits,''
  arXiv:1205.6790 [hep-ph].
  %%CITATION = ARXIV:1205.6790;%%


\bibitem{models} A.~Djouadi,
  %``The Anatomy of electro-weak symmetry breaking. II. The Higgs bosons in the minimal supersymmetric model,''
  Phys.\ Rept.\  {\bf 459}, 1 (2008)
  [hep-ph/0503173]; 
G.~C.~Branco, P.~M.~Ferreira, L.~Lavoura, M.~N.~Rebelo, M.~Sher and J.~P.~Silva,
  %``Theory and phenomenology of two-Higgs-doublet models,''
  Phys.\ Rept.\  {\bf 516}, 1 (2012)
  [arXiv:1106.0034 [hep-ph]].

\bibitem{scalars} V.~Silveira and A.~Zee,
  %``Scalar Phantoms,''
  Phys.\ Lett.\ B {\bf 161}, 136 (1985); T.~Binoth and J.~J.~van der Bij,
  %``Influence of strongly coupled, hidden scalars on Higgs signals,''
  Z.\ Phys.\ C {\bf 75}, 17 (1997)
  [hep-ph/9608245]; J.~McDonald,
  %``Gauge singlet scalars as cold dark matter,''
  Phys.\ Rev.\ D {\bf 50}, 3637 (1994)
  [hep-ph/0702143 [HEP-PH]]; C.~P.~Burgess, M.~Pospelov and T.~ter Veldhuis,
  %``The Minimal model of nonbaryonic dark matter: A Singlet scalar,''
  Nucl.\ Phys.\ B {\bf 619}, 709 (2001)
  [hep-ph/0011335]; D.~O'Connell, M.~J.~Ramsey-Musolf and M.~B.~Wise,
  %``Minimal Extension of the Standard Model Scalar Sector,''
  Phys.\ Rev.\ D {\bf 75}, 037701 (2007)
  [hep-ph/0611014];
V.~Barger, P.~Langacker, M.~McCaskey, M.~J.~Ramsey-Musolf and G.~Shaughnessy,
  %``LHC Phenomenology of an Extended Standard Model with a Real Scalar Singlet,''
  Phys.\ Rev.\ D {\bf 77}, 035005 (2008)
  [arXiv:0706.4311 [hep-ph]].
  
  

\bibitem{impostors}   P.~J.~Fox, D.~Tucker-Smith and N.~Weiner,
  %``Higgs friends and counterfeits at hadron colliders,''
  JHEP {\bf 1106}, 127 (2011)
  [arXiv:1104.5450 [hep-ph]];  R.~Sato, S.~Shirai and T.~T.~Yanagida,
  %``A Scalar Boson as a Messenger of New Physics,''
  Phys.\ Lett.\ B {\bf 704}, 490 (2011)
  [arXiv:1105.0399 [hep-ph]];  I.~Low, J.~Lykken and G.~Shaughnessy,
  %``Singlet scalars as Higgs imposters at the Large Hadron Collider,''
  Phys.\ Rev.\ D {\bf 84}, 035027 (2011)
  [arXiv:1105.4587 [hep-ph]]; I.~Low, J.~Lykken and G.~Shaughnessy,
  %``Have We Observed the Higgs (Imposter)?,''
  arXiv:1207.1093 [hep-ph];
 D.~Bertolini and M.~McCullough,
  %``The Social Higgs,''
  arXiv:1207.4209 [hep-ph].
  %%CITATION = ARXIV:1207.4209;%%


\bibitem{Gunion:2012gc} 
  J.~F.~Gunion, Y.~Jiang and S.~Kraml,
  %``Could two NMSSM Higgs bosons be present near 125 GeV?,''
  arXiv:1207.1545 [hep-ph].

\bibitem{Cline} 
  J.~M.~Cline,
  %``130 GeV dark matter and the Fermi gamma-ray line,''
  arXiv:1205.2688 [hep-ph].
  
  \bibitem{Dobrescu:2000jt} 
  B.~A.~Dobrescu, G.~L.~Landsberg and K.~T.~Matchev,
  %``Higgs boson decays to CP odd scalars at the Tevatron and beyond,''
  Phys.\ Rev.\ D {\bf 63}, 075003 (2001)
  [hep-ph/0005308].
  %%CITATION = HEP-PH/0005308;%%
  
\bibitem{BtoDM} C.~Bird, P.~Jackson, R.~V.~Kowalewski and M.~Pospelov,
  %``Search for dark matter in b ---> s transitions with missing energy,''
  Phys.\ Rev.\ Lett.\  {\bf 93}, 201803 (2004)
  [hep-ph/0401195];  C.~Bird, R.~V.~Kowalewski and M.~Pospelov,
  %``Dark matter pair-production in b ---> s transitions,''
  Mod.\ Phys.\ Lett.\ A {\bf 21}, 457 (2006)
  [hep-ph/0601090].

\bibitem{BABARpaperthatnobodycites} 
  J.~P.~Lees {\it et al.}  [BaBar Collaboration],
  %``Observation of the rare decay $B^+ -> K^+\pi^0\pi^0$ and measurement of the quasi-two body contributions $B^+ -> K^*(892)^+\pi^0$, $B^+ -> f_0(980)K^+$ and $B^+ -> \chi_{c0}K^+$,''
  Phys.\ Rev.\ D {\bf 84}, 092007 (2011)
  [arXiv:1109.0143 [hep-ex]].

\bibitem{minimalmatter}  E.~Del Nobile, R.~Franceschini, D.~Pappadopulo and A.~Strumia,
  %``Minimal Matter at the Large Hadron Collider,''
  Nucl.\ Phys.\ B {\bf 826}, 217 (2010)
  [arXiv:0908.1567 [hep-ph]].
  
  \bibitem{RecentVL}  
  S.~Dawson and E.~Furlan,
  %``A Higgs Conundrum with Vector Fermions,''
  arXiv:1205.4733 [hep-ph];
  %%CITATION = ARXIV:1205.4733;%%
  N.~Bonne and G.~Moreau,
  %``Reproducing the Higgs boson data with vector-like quarks,''
  arXiv:1206.3360 [hep-ph];
  %%CITATION = ARXIV:1206.3360;%%
A.~Joglekar, P.~Schwaller and C.~E.~M.~Wagner,
  %``Dark Matter and Enhanced Higgs to Di-photon Rate from Vector-like Leptons,''
  arXiv:1207.4235 [hep-ph].


\bibitem{oblique} 
    M.~E.~Peskin and T.~Takeuchi,
  %``Estimation of oblique electroweak corrections,''
  Phys.\ Rev.\ D {\bf 46}, 381 (1992);
  %%CITATION = PHRVA,D46,381;%%

\bibitem{zpole} 
  [ALEPH and DELPHI and L3 and OPAL and SLD and LEP Electroweak Working Group and SLD Electroweak Group and SLD Heavy Flavour Group Collaborations],
  %``Precision electroweak measurements on the $Z$ resonance,''
  Phys.\ Rept.\  {\bf 427}, 257 (2006)
  [hep-ex/0509008].
  %%CITATION = HEP-EX/0509008;%%


\bibitem{hfag} 
  Y.~Amhis {\it et al.}  [Heavy Flavor Averaging Group Collaboration],
  %``Averages of b-hadron, c-hadron, and tau-lepton properties as of early 2012,''
  arXiv:1207.1158 [hep-ex].
  %%CITATION = ARXIV:1207.1158;%%


\bibitem{PDG} 
  C.~Amsler {\it et al.}  [Particle Data Group Collaboration],
  %``Review of Particle Physics,''
  Phys.\ Lett.\ B {\bf 667}, 1 (2008).
  %%CITATION = PHLTA,B667,1;%%



\bibitem{Choudhury} 
  D.~Choudhury, T.~M.~P.~Tait and C.~E.~M.~Wagner,
  %``Beautiful mirrors and precision electroweak data,''
  Phys.\ Rev.\ D {\bf 65}, 053002 (2002)
  [hep-ph/0109097].
  %%CITATION = HEP-PH/0109097;%%


\bibitem{AtlasConfNote} ATLAS Collaboration Public Note, Report No. ATLAS-CONF-2012-076, 2012 [\url{http://cdsweb.cern.ch/record/1460273/files/ATLAS-CONF-2012-076.pdf}]. 


  
\bibitem{MadGraph} 
  J.~Alwall, M.~Herquet, F.~Maltoni, O.~Mattelaer and T.~Stelzer,
  %``MadGraph 5 : Going Beyond,''
  JHEP {\bf 1106}, 128 (2011)
  [arXiv:1106.0522 [hep-ph]].
  %%CITATION = ARXIV:1106.0522;%%
  
\bibitem{natural}  N.~Arkani-Hamed, K.~Blum, R.~T.~D'Agnolo and J.~Fan,
  %``2:1 for Naturalness at the LHC?,''
  arXiv:1207.4482 [hep-ph].
  
  \bibitem{4thgenlep} 
  L.~M.~Carpenter, A.~Rajaraman and D.~Whiteson,
  %``Searches for Fourth Generation Charged Leptons,''
  arXiv:1010.1011 [hep-ph].
  %%CITATION = ARXIV:1010.1011;%%
  
\bibitem{Pythia} 
  T.~Sjostrand, S.~Mrenna and P.~Z.~Skands,
  %``PYTHIA 6.4 Physics and Manual,''
  JHEP {\bf 0605}, 026 (2006)
  [hep-ph/0603175].
  %%CITATION = HEP-PH/0603175;%%
  
\bibitem{PGS}
J.~Conway {\it et al.}, \url{http://physics.ucdavis.edu/~conway/research/software/pgs/pgs4-general.htm}.

\bibitem{tdrs}
G.~Aad {\it et al.}  [ATLAS Collaboration],
  %``Expected Performance of the ATLAS Experiment - Detector, Trigger and Physics,''
  arXiv:0901.0512 [hep-ex];
  %%CITATION = ARXIV:0901.0512;%%
G.~L.~Bayatian {\it et al.}  [CMS Collaboration],
  %``CMS technical design report, volume II: Physics performance,''
  J.\ Phys.\ G G {\bf 34}, 995 (2007).
  %%CITATION = JPHGB,G34,995;%%

  


\end{thebibliography}
\end{document}